\documentclass[aps,prd,twocolumn,english,showpacs,superscriptaddress,longbibliography]{revtex4-2}

\usepackage{amsmath}
\usepackage{amssymb}
\usepackage{amsfonts}
\usepackage{mathrsfs}
\usepackage[dvipsnames]{xcolor}

\usepackage{amsthm}

\theoremstyle{definition}

\def\XXint#1#2#3{{\setbox0=\hbox{$#1{#2#3}{\int}$}
     \vcenter{\hbox{$#2#3$}}\kern-.5\wd0}}

\usepackage[utf8]{inputenc}
\usepackage[english]{babel}

\usepackage[]{graphicx}

\usepackage[colorlinks=true,plainpages=false]{hyperref}

\begin{document}

	\title{Analytical control of the exchange interaction in periodically driven Mott insulators}


\author{Miguel Tierz}\email{tierz@simis.cn}
\affiliation{Shanghai Institute for Mathematics and Interdisciplinary Sciences \\ Block A, International Innovation Plaza, No. 657 Songhu Road, Yangpu District,\\ Shanghai, China}
\affiliation{Departamento de An\'alisis Matem\'atico y Matem\'atica Aplicada, Universidad Complutense de Madrid, 28040 Madrid, Spain}
 
 \begin{abstract}
The manipulation of electronic structure through periodic electric fields enables the reversible control of effective interactions in extended antiferromagnetic Mott insulators on ultrafast timescales. A careful analytical examination of the modulated effective interactions is conducted, accurately characterising it through the use of exact summation formulas and Bessel functions. As a result, time reversals are analytically determined in terms of Bessel zeroes. We discuss the half-filled Hubbard model, as well as multi-orbital models, various characteristics of the Kitaev-Heisenberg model, and the emergence of chiral spin terms.
\end{abstract}


\maketitle  The manipulation and control of many-body states through their coupling to
external degrees of freedom have garnered significant attention and success in recent years \cite{bukov2015universal,oka2019floquet,de2021colloquium,rudner2020band}. Equilibrium macroscopic phenomena
are described by thermodynamics, relying only on a small number of state
variables such as temperature and pressure. However, when a system is driven
continuously by external means, a plethora of novel
phenomena emerges. In particular, under the application of light periodic in
time a quantum system can be described with a time independent effective
Hamiltonian, which is known as Floquet Hamiltonian \cite{floquet1883equations,shirley1965solution,rodriguez2021low}. 

The Floquet
theorem implies that the time evolution in steps of the driving period $T$ can
be described by a time-independent Hamiltonian. Significant interest in Floquet quantum systems has been spurred by recent experimental advancements in fields such as cold atoms and trapped ions, which have enabled the realization of coherent Floquet quantum dynamics in many-body systems.

From the aforementioned concepts, Floquet engineering \cite{oka2019floquet} has emerged as a field that involves controlling many-body states using periodically driven external light.
 
This external field has proven to be an efficient tool for adjusting magnetic interactions \cite{bukov2016schrieffer,mentink2015ultrafast,mentink2017manipulating}. For example, by applying $E(t)=E_{0}\cos\left(  \omega t\right)  $ to
a single-orbital Mott insulator and tuning $E_{0}$ and $\omega$, the magnitude
and sign of the antiferromagnetic Heisenberg interaction between magnetic
moments can be changed. We will emphasize the analytical study of the reversals first discussed in \cite{mentink2015ultrafast}.

The Hubbard model is given by the Hamiltonian \cite{hubbard1964electron,arovas2022hubbard}
\begin{equation}
H=-t_{0}\sum_{\langle ij\rangle\sigma}c_{i\sigma}^{\dagger}c_{j\sigma}%
+U\sum_{j}n_{j\uparrow}n_{j\downarrow} \label{repulsive hubbard}%
\end{equation} where $c_{i\sigma}^{\dagger}$ creates an electron at site $i$ with spin
$\sigma=\uparrow,\downarrow$, $t_{0}$ is the hopping between nearest-neighbor
sites and $U$ the repulsive on-site interaction $U$. It describes a
Mott-insulator with one electron per site for half-filling and at $U/t_{0}%
\gg1$ \cite{arovas2022hubbard}. Because of the Pauli principle, hoppings are possible when adjacent
sites have opposite spin. It is convenient to include also a homogeneous
static magnetic field given by%
\begin{equation}
H_{Z}=B_{x}\sum_{j}S_{jx}.
\end{equation}
where the spin $S_{j\alpha}=\frac{1}{2}\sum_{\sigma\sigma^{\prime}}c_{j\sigma
}^{\dagger}(\hat{\sigma}_{\alpha})_{\sigma\sigma^{\prime}}c_{j\sigma^{\prime}%
}$ is coupled to a homogeneous magnetic field $B_{x}$ along the $x$ axis
($\alpha=x,y,z$; $\hat{\sigma}_{\alpha}$ denote the Pauli matrices). 

Non-equilibrium dynamics resulting from time-dependent electric fields $\mathbf{E}(t)$ can
be most efficiently incorporated by adding a Peierls phase \cite{peierls1933theorie,luttinger1951effect}
to the hopping process:
\begin{equation}
t_{ij}(t)=t_{0}\exp\left[ {\mathrm{i}e\mathbf{A}(t)\cdot(\mathbf{R}_{i}-\mathbf{R}_{j})/\hbar}\right] ,
\end{equation}
The homogeneous vector potential
is expressed as $\mathbf{A}(t)=-\partial_{t}\mathbf{E}(t)$. This can be understood in terms of a gauge where a
scalar potential is added at each site, denoted by $e\phi_{i}(t)\sim
\mathbf{R}_{i}\cdot\mathbf{E}(t)$. As previously stated, Floquet's theorem \cite{floquet1883equations,shirley1965solution} can be applied in the presence of electric fields that vary periodically with time. In
Floquet theory, the solutions of the time-dependent Schr\"{o}dinger equation
take the form $|\psi(t)\rangle=e^{-\text{i}\epsilon_{\alpha}t}%
|\psi_{\alpha}(t)\rangle$ where $|\psi_{\alpha}(t+T)\rangle=|\psi_{\alpha
}(t)\rangle$ is time-periodic with a period $T=2\pi/\omega$, and
$\epsilon_{\alpha}$ is a quasi-energy defined up to multiples of $\omega$. Alongside the virtual hoppings responsible for determining equilibrium $J_{\mathrm{ex}}$, there is also virtual absorption and emission of photons into differing Floquet
sectors. This
mixing between sectors has a critical role
in renormalizing the quasi-energy levels. 

For half-filling and $U/t_{0}\gg 1$ the Hubbard model describes a
Mott insulator with one electron per site, in which the remaining, low-energy,
spin degrees of freedom are coupled by an effective antiferromagnetic Heisenberg exchange
interaction $J_{\text{ex}}$ \cite{mentink2015ultrafast,auerbach1998interacting}.  

Floquet engineering is also undertaken in the exploration of potential Kitaev-Heisenberg materials \cite{kitaev2006anyons} that are expected to contain a Kitaev spin liquid \cite{savary2016quantum,broholm2020quantum,motome2020hunting,hermanns2018physics,trebst2022kitaev}. Driving with time-periodic light enables the adjustment of the relative strength of interactions, an area of interest when it comes to applying it to materials like iridates and the ruthenate $\alpha $-RuCl$_{3}$, which are already in close proximity to a Kitaev phase. 

First, for the single-band Hubbard model, the amplitude $E_{0}$ of the external field ends up appearing
as the dimensionless driving strength $\mathcal{E}=eaE_{0}/\hbar\omega$, where
$e$ and $a$ are unit charge and lattice spacing, respectively. 

The corresponding effective Hamiltonian was obtained for the case $\Omega ,U\gg v_{0}$: For off-resonant driving ($U\neq l\Omega $), no doublons (D) and
holons (H) are generated, and the leading term of the effective Hamiltonian becomes a spin Hamiltonian
with renormalized exchange coupling      
\begin{equation}
J_{\text{ex}}(\mathcal{E},\omega)=\sum_{m=-\infty}^{\infty}\frac{2t_{0}%
^{2}J_{|m|}(\mathcal{E})^{2}}{U+m\hbar\omega}, \label{Jeff 1}%
\end{equation}
where $J_{m}(\mathcal{E})$ denote the usual Bessel function of the first type. This expression is now central in the field and shows that the exchange coupling includes the contributions of virtual creations of DH pairs dressed with $l$ photons.

The corresponding Floquet spin Hamiltonian was obtained using the
Floquet Schrieffer-Wolff transformation \cite{bukov2016schrieffer}.
J. Mentink et al. \cite{mentink2015ultrafast} used perturbation theory in $v_{0}/U$ in the extended Floquet
Hilbert space, while \cite{kitamura2017probing} employed a time-dependent Schrieffer-Wolff transformation, \cite{itin2015effective} a variant of the high-frequency
expansion and \cite{hejazi2019floquet} introduced it developing a time- dependent perturbation theory.  

Therefore, there are amplitude and frequency ranges where the exchange coupling becomes ferromagnetic (FM \cite{mentink2015ultrafast}. A sign-change of $J_{\text{ex}}$ in these ranges will not induce a transition to a FM state since the Hubbard model Hamiltonian conserves the total spin \cite{mentink2015ultrafast}. Nonetheless, a periodic driving-induced change in the sign of $J_{\text{ex}}$ presents a distinctive means to manipulate the spin dynamics, i.e. the reversal of the time evolution of the system. There is extensive numerical analysis of this, but no analytical treatment, which we provide in the following.

 Commencing with an assessment of the Bessel summation, the reversal can be accurately and thoroughly quantified analytically \footnote{It is widely recognised that this expression is frequently encountered in the investigation of quantum Floquet systems - and beyond.}. The analytical evaluation of \eqref{Jeff 1} \cite{sen1952solar,newberger1982new} remains unused \cite{russo2024landau}
\begin{equation}
J_{\text{ex}}(\mathcal{E},\omega)=\frac{2t_{0}^{2}\pi}{\hbar\omega\sin\left(
\pi\mu\right)  }J_{\mu}(\mathcal{E})J_{-\mu}(\mathcal{E}), \label{Jeff 2}%
\end{equation}
where $\mu\equiv U/\hbar\omega$. The physical resonances
\cite{bukov2016schrieffer,mentink2015ultrafast} correspond to the non-integer restriction in the summation formula, given by $\mu \in \mathbb{C}/\mathbb{Z}$ \cite{newberger1982new,russo2024landau}. The expression exhibits manifest simplicity and symmetry, brought about by the final two Bessel functions with an order of $\mu$. From this expression, it is immediate to obtain limiting behavior. For $\varepsilon \gg \frac{U^{2}}{\omega ^{2}}$, we find: 
\begin{equation}
J_{\text{ex}}(\mathcal{E},\omega )\simeq \frac{2t_{0}^{2}}{\hbar \omega
\varepsilon }\frac{\cos \left( \pi \mu \right) +\sin (2\varepsilon )}{\sin
(\pi \mu )},  \label{Jeff 2}
\end{equation}%
Corrections to this formula can be obtained, extending the
validity, say, to $\varepsilon \gg \frac{\left\vert U\right\vert }{\omega }$. 
For small amplitudes $\varepsilon \rightarrow 0$ \cite{russo2024landau}
\[
J_{\text{ex}}(\mathcal{E},\omega )\simeq \frac{2t_{0}^{2}}{\hbar \omega \mu }%
\left( 1+\frac{\varepsilon ^{2}}{2(\mu ^{2}-1)}\right) .
\]
In the supplementary material we provide mathematical details and extensions. Moving forward, we exploit \eqref{Jeff 2} to provide an analytical description of the sign reversals.

After discussing the Hubbard model and its time evolution reversals \cite{mentink2015ultrafast,mentink2017manipulating} we will extend the discussion to the case of multi-orbital models, Kitaev magnets and the emergence of spin chiral terms \cite{kitamura2020current}.
\section{Sign reversals of the exchange interaction at the Bessel function
zeros}

The search for the solution of mathematical zeros of Bessel functions has been ongoing for over a century.  Typically, one fixes the function, which in this instance is the order, corresponding to the Hamiltonian parameter U, and the values of the external force amplitude that produce zero values are explored. However, another approach is to consider a fixed argument
$\mathcal{E}_{0}$ and explore $J_{\mu}(\mathcal{E}_{0})$ as a
function of $\mu$. This
corresponds to adjusting the driving and searching for the values of U that result in
sign reversals. 
According to \cite{coulomb1936zeros} there are an infinite number of such values that are well-spaced apart.

\subsection{Maximal distance between resonances}

If we set $U$ to be precisely symmetrically located between the resonance values. That is, at half-integer values of $U$, the expression becomes even simpler and has special properties. The formula corresponds then to the spherical Bessel functions which consists of trigonometric functions and monomials. We consider the two simplest cases
\begin{itemize}
    \item For $U=\pm\frac{m\omega}{2}$
    \begin{equation}
J_{\text{ex}}(\mathcal{E},\omega)=\frac{2t_{0}^{2}\pi}{\hbar\omega}%
J_{1/2}(\mathcal{E})J_{-1/2}(\mathcal{E})=\frac{2t_{0}^{2}}{\hbar\omega}%
\frac{\sin2\mathcal{E}}{\mathcal{E}},\label{one half}%
\end{equation}
\item For $U=\pm\frac{3m\omega}{2}$ \footnote{Notice how the $3/2$ case can be written also as the $1/2$
case plus extra terms subleading for large external driving.} 
\begin{equation}
J_{\text{ex}}(\mathcal{E},\omega)=\frac{2t_{0}^{2}}{\hbar\omega}\frac
{1}{\mathcal{E}}\left(  \left(  1-\frac{1}{\mathcal{E}^{2}}\right)
\sin2\mathcal{E}+\frac{\cos^{2}\mathcal{E}-\sin^{2}\mathcal{E}}{2\mathcal{E}%
}\right)  .\nonumber
\end{equation}
\end{itemize}
The case \eqref{one half} is the only purely trigonometric and hence the only
one with equispaced sign reversals, for values of the driving $\mathcal{E}%
_{\mathrm{crit}}=n\pi/2$ for $n\in\mathbb{Z}.$ In figure 1 we see the behavior.

\begin{figure}[!htb]
  \includegraphics[width=\linewidth]{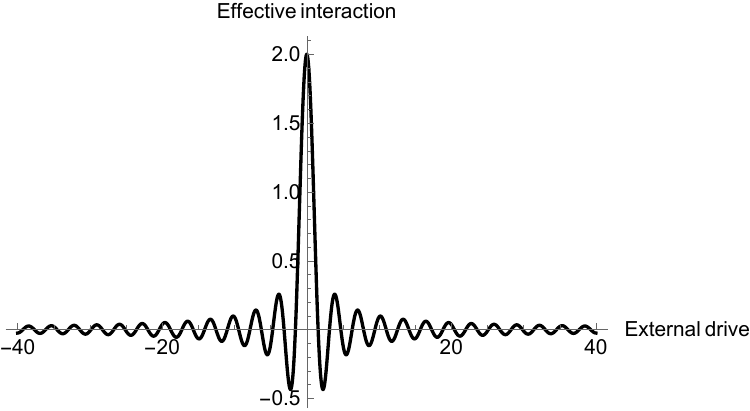}
  \caption{$\mu=0.5$}\label{fig:awesome_image1}
\end{figure}
As a matter of fact, from Sturm's comparison theorem, it follows that the
zeros of $J_{\nu}(x)$ \footnote{More, generally, the zeros of the cylindrical function,
hence the result also applies to $J_{-\nu}(x)$ among others.} satisfy that, if $j_{\nu,s}$ and  $j_{\nu,s1}$ denote
two positive zeros of $J_{\nu}(x)\ $then%
\begin{align*}
|\nu|  & \neq1/2\rightarrow\text{ }(j_{\nu,s1}-j_{\nu,s}-\pi )(|\nu
|-1/2)>0.\\
|\nu|  & =1/2\rightarrow(j_{\nu,s1}-j_{\nu,s}-\pi )=0.
\end{align*}
Physically, when $U/\omega$ is less than half-resonance, the spacing between reversals (zeros)
is smaller than $\pi$ and increases as it approaches half-resonance.
At that point, it is exactly $\pi$ and afterwards, the spacing steadily decreases, resulting
in more frequent reversals. In other words, the function $j_{\nu,s}$ is concave in relation
to $s$ when $|\nu |>1/2$ and convex when $0\leq|\nu |<1/2$. 

It is important to note that if we initiate the process with a small positive value of $U /\omega$ and we are less than halfway to a resonance, there is a significant bias towards positive values of $J_{\text{ex}}(\mathcal{E},\omega)$ and the reversal to a negative value is of small amplitude. After passing the half-integer point, the opposite is observed until we reach a resonance, at which point the cycle repeats itself (see Figure 2).

It is manifest from Figure 3 that the behavior of $J_{\text{ex}}(\mathcal{E},\omega)$ differs significantly just below a resonance, where it is strongly negative, and immediately after passing the resonance, where it is mostly positive. Figure 3 demonstrates that both the above and below resonance points show a very dominant sign of the effective interaction. 

Note how Figure 3 delineates the proximity to resonance from both the below and above perspectives exhibiting, with the exception of small values of the drive, a near exact mirror-image relationship with respect to the $\mathcal{E}$ axis. It is also noteworthy from Figure 3 that,  in the vicinity of the resonance, both above and below it, the minoritary sign region is minuscule, though not identically zero.

\begin{figure}[!htb]
  \includegraphics[width=\linewidth]{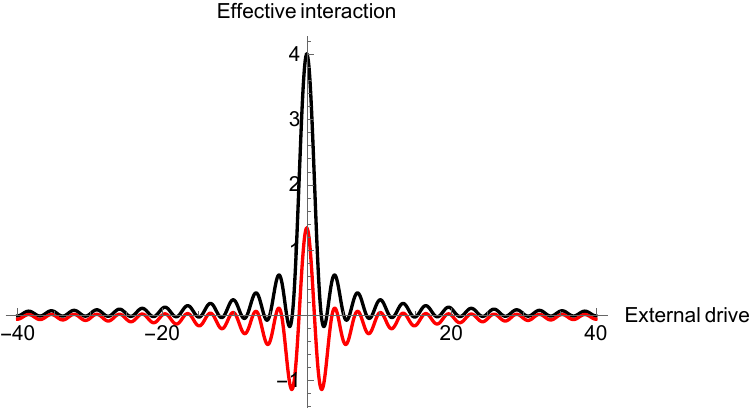}
  \caption{$\mu=0.25$ (Black) and $\mu=0.75$ (Red)}\label{fig:awesome_image2}
  \includegraphics[width=\linewidth]{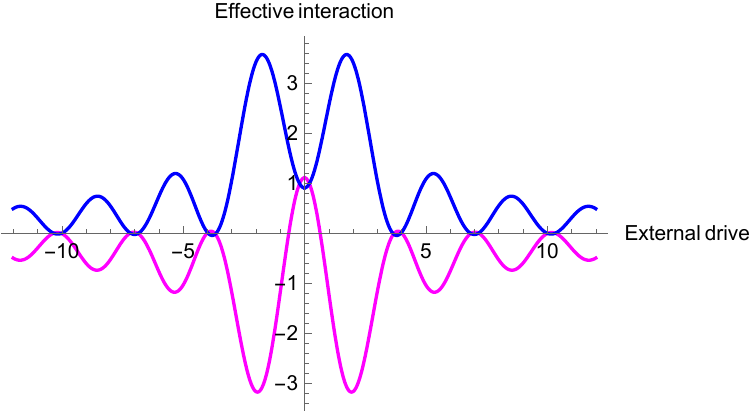}
  \caption{$\mu=0.9$ (Magenta) and $\mu=1.1$ (Blue) }\label{fig:awesome_image3}
\end{figure}  



\subsection{Fixing drive $\mathcal{E}$ and tuning $U$: Super-exponential proximity of time reversals to resonances}

The mathematical work \cite{coulomb1936zeros} examined the behavior of the zeros if the variable is held constant
and the order of the Bessel function varied. Since $U$ is a tunable parameter of the Hubbard model for example in cold atom simulations, this setting is physical. This has the main feature of encountering the infinite
number of resonances. 

 It has been observed that as $\mu\rightarrow-\infty$, the zeros converge to the negative integers
\cite{coulomb1936zeros}. This was made precise in \cite{flajolet1990non}, where it was proven that,
fixing what is the external drive at $\mathcal{E}\mathbb{=}2$, it holds that
the $rth$ negative zero $\nu_{r}$ of $J_{\nu}\left(  \mathcal{E}%
=2\right)$ satisfies
\[
\nu_{r}=-r+\frac{1}{r!(r-1)!}+O\left(  \frac{1}{\left(  r!\right)  ^{2}%
}\right) \label{eq:fund} ,
\]
and hence most of the points of time reversal are \emph{super exponentially close} to the resonance condition. See Figures 1, 2 and 3 to see this delicate
situation, due to the near coincidence, when fixing the external drive and tuning the $U$ parameter, between the zeros of the Bessel function
and the poles of the formula \eqref{Jeff 2}, describing the resonances given by the zeros of
the sine function. 


\begin{figure}[h] 
\includegraphics[width=0.37\textwidth]{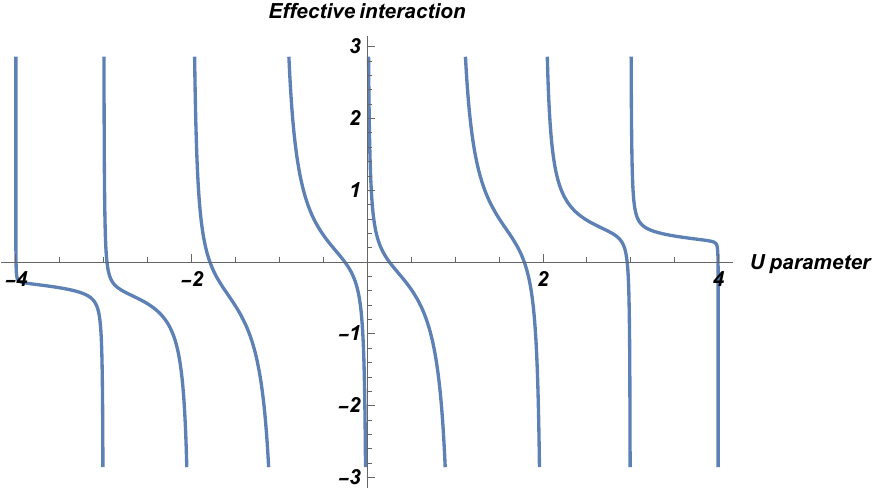}
\caption{The effective interaction $J_{\text{ex}}(\mathcal{E}=2,\omega=1)$ as a function of $U$.}
\end{figure}

\begin{figure}[h] 
\includegraphics[width=0.37\textwidth]{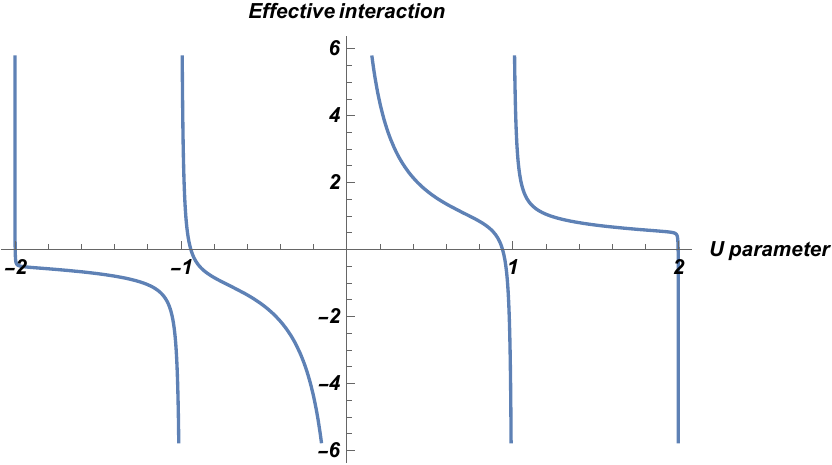}
\caption{Effective interaction $J_{\text{ex}}(\mathcal{E}=0.5,\omega=1)$}
\end{figure}



\begin{figure}[h] 
\includegraphics[width=0.37\textwidth]{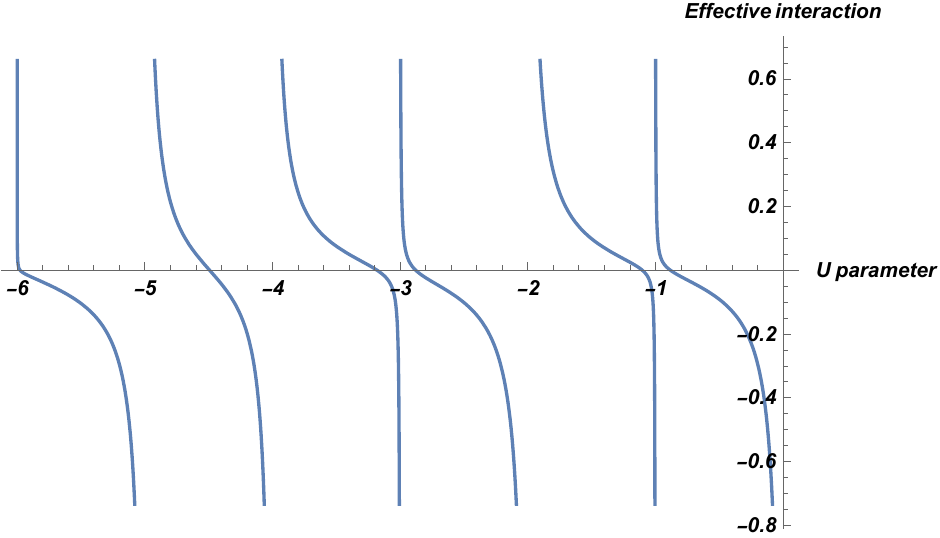}
\caption{Effective interaction $J_{\text{ex}}(\mathcal{E}=10,\omega=1)$}
\end{figure}

For $\mathcal{E}\mathbb{=}2$ only the first two zeros (per side, $4$ in total then) are well-separated, as can be seen in Figure 1 (see numerical values in \cite{flajolet1990non}). One result of this phenomenon is the presence of minute reversals before we experience a reset following a resonance.

The greater the value of $U$, the more significant this effect becomes.
 In the case of $\mathcal{E}=2$ for example, there remain one or two reversals (both positive and negative \footnote{Recall the symmetry of the
formula. Hence, the reversals are near positive and negative integers.}) which
are distinctly separated from the resonance. 

Numerical evaluations 
reveal that for smaller external drives, like $\mathcal{E}\mathbb{\leq}1$,
all reversal $U$ values are in close proximity to a resonance, as illustrated in Figure 2. Conversely, for larger drives, such as $\mathcal{E}\mathbb{=}10$ or
$\mathcal{E}\mathbb{=}20$ we have that the highly regular pattern above
is disturbed for a larger number of the first zeros (see Figure 3). Therefore we have more
reversals which are well separated from the resonance.  

\section{Multi orbital models}

The analysis of periodically driven Hubbard models has been extended
to several other models. For example, to periodically driven multi-orbital
Mott insulators, characterized by Kanamori local interactions
\cite{kanamori1959superexchange}. The Hubbard-Kanamori model expands upon the Hubbard model to include many bands and consider the Hund coupling.
\begin{align}
H_{\text{K}}= &  U\,\sum_{i,\alpha}\hat{n}_{i\alpha\uparrow}\hat{n}%
_{i\alpha\downarrow}+U^{\prime}\sum_{i,\alpha<\beta,\sigma,\sigma^{\prime}%
}\hat{n}_{i\alpha\sigma}\hat{n}_{i\beta\sigma^{\prime}}\;\nonumber\\
&  -J_{\text{H}}\sum_{i,\alpha<\beta,\sigma,\sigma^{\prime}}c_{i\alpha\sigma
}^{\dagger}c_{i\alpha\sigma^{\prime}}c_{i\beta\sigma^{\prime}}^{\dagger
}c_{i\beta\sigma}\;\nonumber\\
&  +J_{\text{P}}\sum_{i,\alpha<\beta,\sigma}c_{i\alpha\sigma}^{\dagger
}c_{i\alpha-\sigma}^{\dagger}c_{i\beta\sigma}c_{i\beta-\sigma}%
\;,\label{eq:kanamori}%
\end{align}
with $U$ and $U^{\prime}$ the intra-orbital and inter-orbital direct Coulomb
interactions. $J_{\text{H}}$ refer to the onsite exchange interaction, known
as Hunds'coupling \cite{georges2013strong} and $J_{\text{P}}$ to the pair
hopping. The sets of indices $\{i,j\}$, $\{\alpha,\beta\}$, $\{\sigma
,\sigma^{\prime}\}$ denote the lattice sites, orbitals and spin degrees of
freedom. 

Its application lies in characterizing crystals with d- and f-shells that are partly filled and demonstrate moderate to strong on-site Coulomb repulsion.

The effect of the external drive is
manifested in the kinetic energy $T_{t}$ once more, through the Peierls
substitution, whereby the expression for the hopping amplitudes $t_{i\alpha,j\beta}$ conforming $T_{t}$ is extended to include the dependence on orbitals.
$t_{i\alpha,j\beta}$ represents the hopping amplitude from orbital $\beta$ at
site $j$ to orbital $\alpha$ at site $i$ but the Peierls phase factor is of
the same type as above. It is then obtained that \cite{liu2018floquet,hejazi2019floquet} \
\begin{align*}
\hat{J}_{ij}= &  \sum_{n=-\infty}^{\infty}J_{n}^{2}(u_{ij})\Bigg\{\frac
{\hat{\gamma}_{ij,2}-\hat{\gamma}_{ij,1}}{U^{\prime}-J_{\text{H}}-n\omega
}+\frac{\hat{\gamma}_{ij,2}+\hat{\gamma}_{ij,1}-\hat{\gamma}_{ij,3}}%
{U^{\prime}+J_{\text{H}}-n\omega}\\
&  \qquad\qquad+\frac{\hat{\gamma}_{ij,3}-\hat{\gamma}_{ij,4}}{U-J_{\text{P}%
}-n\omega}+\frac{\hat{\gamma}_{ij,4}}{U+2J_{\text{P}}-n\omega}\Bigg\}.
\end{align*}
The $\hat{\gamma}_{ij,k}$ with $k=1,2,3,4$ are orbital operators and hence the
effective exchange $\hat{J}_{ij}$ denotes now an orbital operator. Following
as above, this means that this effective operator can be written as
\[
\hat{J}_{ij}=\frac{1}{\omega}\sum_{i=1}^{4}\frac{\pi\widehat{a}_{i}}%
{\sin\left(  \pi\mu_{i}\right)  }J_{\mu_{i}}(\mathcal{E})J_{-\mu_{i}%
}(\mathcal{E}),
\]
with $\widehat{a}_{i}$ denoting the orbital operators above after trivial
identification and $\mu_{1}=$ $\left(  J_{\text{H}}-U^{\prime}\right)
/\omega$, $\mu_{2}=$ $\left(  -J_{\text{H}}-U^{\prime}\right)  /\omega$,
$\mu_{3}=$ $\left(  -J_{\text{P}}-U\right)  /\omega$ and $\mu_{4}=$ $\left(
-2J_{\text{P}}-U\right)  /\omega$. 

There are two settings that can result in reversals.
 The first happens when the total sum equals zero.
The second occurs when all terms in a sum, such as this one,
are identical and equal to zero,
referred to as shared roots of two or more distinct
Bessel functions.

We analyze the case corresponding to the
Kanamori Hamiltonian being rotational invariant, which happens for $U^{\prime
}=U-J_{H}$ and $J_{P}=0$ \cite{georges2013strong}. Then, the operator becomes, after employing the Bessel summation formula \cite{newberger1982new,russo2024landau},
and reorganizing the expression in terms of the orbital operators%
\begin{align}
\frac{\omega}{\pi}\hat{J}_{ij}  & =\hat{\gamma}_{ij,1}\left(  \frac
{J_{\lambda_{2}}(\mathcal{E})J_{-\lambda_{2}}(\mathcal{E})}{\sin\left(
\pi\lambda_{2}\right)  }-\frac{J_{\lambda_{1}}(\mathcal{E})J_{-\lambda_{1}%
}(\mathcal{E})}{\sin\left(  \pi\lambda_{1}\right)  }\right)  \\
& +\hat{\gamma}_{ij,2}\left(  \frac{J_{\lambda_{2}}(\mathcal{E})J_{-\lambda
_{2}}(\mathcal{E})}{\sin\left(  \pi\lambda_{2}\right)  }+\frac{J_{\lambda_{1}%
}(\mathcal{E})J_{-\lambda_{1}}(\mathcal{E})}{\sin\left(  \pi\lambda
_{1}\right)  }\right)
\end{align}
with $\lambda_{1}=$ $\left(  2J_{\text{H}}-U\right)  /\omega$ and $\lambda
_{2}=-U/\omega$. We can now examine reversals at this operator level. If
we wish to set the entire $\hat{J}_{ij}$ to be $0$, it is evident from the
equation that consideration of common zeros are advantageous in this case, as each term must equal $0$.

As a positive result, regarding the existence of common zeros, it holds that,
when $\rho$ is real, there exists a sequence of real numbers $\left\vert
\nu_{m}\right\vert \rightarrow\infty$ $m\rightarrow\infty$, such that
$J_{\nu_{m}}\left(  \rho\right)  =0.$ This guarantees that there are reversal
points for $\hat{J}_{ij}$ by choosing, after fixing the drive $\mathcal{E}$,
$\lambda_{1}$ and $\lambda_{2}$ as two members of such a sequence. 

It is simpler to interpret reversal of effective exchange interactions,
which, under the condition that $\left\vert U-n\omega \right\vert $ and/or
$\left\vert U-2JH-n\omega \right\vert $ is much greater than the typical
hopping amplitudes and, after averaging the orbital operators, becomes,
after using the summation formula:%
\begin{align*}
\overline{J}_{ij}  & =\frac{\pi\left(  \gamma_{1}+\gamma_{2}\right)  }%
{\omega\sin\left(  \pi U/\omega\right)  }J_{U/\omega}(\mathcal{E}%
)J_{-U/\omega}(\mathcal{E})\\
& +\frac{\pi\left(  \gamma_{1}-\gamma_{2}\right)  J_{(U-2J_{H})/\omega
}(\mathcal{E})J_{-(U-2J_{H})/\omega}(\mathcal{E})}{\omega\sin\left(
\pi(U-2J_{H})/\omega\right)  },
\end{align*}
where the first term is the antiferromagnetic component from all of the single
virtual excitations and the second one the ferromagnetic component from all of
the triplet virtual excitations \cite{liu2018floquet}. Fixing $U$, there is an infinite number of zeros, moving the driving parameter. The asymptotics for large $\mathcal{E}$ can now be given: 
\begin{align*}
  \overline{J}_{ij}  & \approx \frac{\left( \gamma _{1}+\gamma _{2}\right) }{\mathcal{E}\omega }\frac{\cos \left( \frac{\pi U}{\omega }\right) +\sin \left(
2\mathcal{E} \right) }{\sin \left( \frac{\pi U}{\omega }\right) }\\
& +\frac{\left( \gamma _{1}-\gamma _{2}\right) }{\mathcal{E}\omega }\frac{\cos \left( \frac{\pi
(U-2J_{H})}{\omega }\right) +\sin \left( 2\mathcal{E} \right) }{\sin \left( 
\frac{\pi (U-2J_{H})}{\omega }\right) }.
\end{align*}
Fine tuning the frequencies such that $\frac{2J_{H}}{\omega }=2n-1$ or $%
\frac{2J_{H}}{\omega }=2n$ for $n\in \mathbb{Z}
$ we have the following simplifications. For even integer:
\begin{equation*}
  \overline{J}_{ij} \approx \frac{2\gamma _{1}}{\omega \mathcal{E} }\left( \tan \left( \frac{\pi U}{%
\omega }\right) +\frac{\sin \left( 2\mathcal{E} \right) }{\sin \left( \frac{%
\pi U}{\omega }\right) }\right) ,  
\end{equation*}
and is independent of $\gamma _{2}$. For odd integer:
\begin{equation*}
    \overline{J}_{ij}  \approx
\frac{1}{\omega \mathcal{E} }\left( 2\gamma _{2}\tan \left( \frac{\pi U}{%
\omega }\right) +\gamma _{1}\frac{\sin \left( 2\mathcal{E} \right) }{\sin
\left( \frac{\pi U}{\omega }\right) }\right) .
\end{equation*}
In both cases, (trigonometric) zeroes can be quickly found. 

\subsection{Other models and terms: Kitaev quantum magnets and spin chiral terms}

The Kitaev honeycomb model \cite{kitaev2006anyons} and its extensions to other tricoordinated lattices represent the principal theoretical models where the ground state is recognized as a quantum spin liquid \cite{savary2016quantum,broholm2020quantum}. Quantum spin liquids have been of sustained interest due to their unique properties of spin fractionalization and long-range entanglement \cite{savary2016quantum,broholm2020quantum,motome2020hunting}. The Kitaev interaction could be the preeminent spin interaction in specific transition metal compounds, such as iridates, with robust spin-orbit coupling \cite{khaliullin2005orbital,jackeli2009mott}. This discovery has led to the Kitaev materials field \cite{winter2016challenges,hermanns2018physics,trebst2022kitaev}.  

In \cite{kumar2022floquet}, they investigate the widely recognised $J-K-\Gamma $ spin-exchange model \cite{rau2014generic,rau2016spin} with external
driving, while considering the ligand degrees of freedom by utilising third-order perturbation theory.

In the absence of ligand effects, the model comprises of three terms: the effective coupling $J$, the Kitaev term $K$, the Gamma term $\Gamma$ and anisotropy $\Gamma$. These terms are easily expressed following the multi-orbital discussion above, as it only involves the standard Bessel summation formula, the one leading to \eqref{Jeff 2} \footnote{In the Appendix, we provide a discussion of the
complete setup.}. 

However, incorporating ligand effects necessitates double nested infinite summations, which can still be achieved by iteratively utilising Eq.  \eqref{Jeff 2}.
This is achieved provided we slightly simplify
the setup described in \cite{kumar2022floquet}. 

For definiteness, we focus on the result \cite{kumar2022floquet} for the the Kitaev term $K$ and the other two follow in the same manner. In \cite{kumar2022floquet} they obtain
\begin{align*}
      K & = \frac{8 t_{pd}^2}{9} \sum_{n,m = -\infty}^{\infty} \mathfrak{J}_{m,n} (\zeta) \bigg[ \frac{J_{\mathrm{H}}}{(\Delta + m \omega)} \times \\
& \frac{\sin[(m-n)\psi_0] (t_1-t_3) -3\cos[(m-n)\psi_0]t_2}{(E_P+(m+n)\omega)(E_D+(m+n)\omega)}\bigg], 
\end{align*}
where $\mathfrak{J}_{m,n} (\zeta)=\mathcal{J}_{m+n}(\zeta )\mathcal{J}_{-m}(\zeta
r_{ij}/R_{ij})\mathcal{J}_{-n}(\zeta r_{ij}/R_{ij})$ and $r_{ij}$ is the
bond-length between the transition metal (TM) and ligand sites, whereas $\psi
_{0}$ is the angle between the TM-TM and TM-ligand bond. Using the following two approximations, with regards to \cite{kumar2022floquet},
\begin{itemize}
    \item To take $r_{ij}/R_{ij}=1$ in order to be able to use \eqref{Jeff 2} iteratively.
    \item We simplified also to $\psi _{0}=0$ 
    \end{itemize}

In the supplementary material (Section C and D, respectively) we discuss in more detail partial extensions of Bessel summation with the aim of avoiding the need for such simplifications. After these two simplifications, one obtains
\begin{align}
K& =\frac{-24t_{pd}^{2}J_{\mathrm{H}}t_{2}J_{\frac{\Delta }{\omega }}(\zeta )%
}{9\omega ^{2}\sin \left( \pi \Delta \right) \left( E_{D}-E_{P}\right) }%
\times  \\
& \left( \frac{J_{-\frac{E_{P}}{\omega }}(\zeta )J_{\frac{E_{P}-\Delta }{%
\omega }}(\zeta )}{\sin \left( \frac{\pi E_{P}}{\omega }\right) }-\frac{J_{-%
\frac{E_{D}}{\omega }}(\zeta )J_{\frac{E_{D}-\Delta }{\omega }}(\zeta )}{%
\sin \left( \frac{\pi E_{D}}{\omega }\right) }\right) ,  \nonumber
\end{align}%

Other multiple summations that can be carried out explicitly are those of \cite{chaudhary2019orbital,chaudhary2020controlling},
also including ligand effects.

\vspace{0.5cm}

\subsubsection{Spin chiral terms}



Beyond the use of linearly polarised light, the application of circularly polarised light to frustrated lattices, including triangular or honeycomb lattices, produces a scalar spin chiral term proportional to $\mathbf{S}_{i}\cdot (\mathbf{S}_{j}\times \mathbf{S}_{k})$, which breaks time reversal symmetry while preserving $SU(2)$ symmetry \cite{kitamura2017probing, oka2019floquet, sur2022driven}. 


In \cite{kitamura2017probing}, in their study of emerging scalar spin chiral terms, they obtain an effective static Hamiltonian as a Heisenberg model that has an emergent scalar chirality term which is of the form
\begin{equation}
J_{\chi ,ijk}^{\text{(h)}}=-A\sum_{m=1}^{\infty }a_{m}(\alpha _{ij},\alpha
_{jk},\frac{U}{\omega })\sin 2m(\theta _{ij}-\theta _{jk}),  \label{scalar}
\end{equation}
where the Fourier coefficients $a_{m}(\alpha _{ij},\alpha _{jk},\frac{U}{\omega })$
are given by a complicated double Bessel sum, that we write down and we evaluate, by iterated use of \eqref{general}, in the supplementary material. We have twice the number of parameters, namely $\alpha _{ij}$ and $\alpha_{jk}$ and $\theta _{ij}$ and $\theta _{jk}$ because this term emerges at fourth-order perturbation theory. Again, the oscillating electric field is described by the Peierls phase $A_{ij}(t)$ (from site $i$ to site $j$) and a monochromatic laser was considered $A_{ij}(t)=A_{ij}\cos({\Omega t-\theta _{ij}})$ \cite{kitamura2017probing, oka2019floquet}.
\\ 

Here, after explicitly obtaining the $a_{m}$, we give their large $\alpha _{ij},\alpha _{jk}$ expansion, which simplifies considerably and, due to the periodicity of the trigonometric functions and the basic form \eqref{Wlargex}, is independent of $m$ \footnote{The opposite limit is also immediate but that one does not require the full summation and is already given in \cite{kitamura2017probing}}
\begin{equation}
a_{m}\sim \frac{4\left( \cos \left( \frac{\pi U}{\omega }\right) +\sin
2\alpha _{ij}\right) ^{2}\left( \cos \left( \frac{\pi U}{\omega }\right)
+\sin 2\alpha _{jk}\right) ^{2}}{m\sin ^{4}\left( \frac{\pi U}{\omega }%
\right) }  \label{largeas}
\end{equation}
In fact, the only dependence on $m$ is the one already explicit in \eqref{scalar} (see \eqref{eq:eff-chiral-1}). There is actually an $m$-dependence encoded in the validity region for the above asymptotic result,
which is given by $\left\vert \alpha _{ij}\right\vert ,\left\vert \alpha
_{jk}\right\vert \gg \left( \frac{U}{\omega }+2m\right) ^{2}$. Thus, the higher the order of the Fourier coefficient the lower the domain of validity of the expansion due to the ensuing increase in the value of $%
\left\vert \alpha _{ij}\right\vert $ and $\left\vert \alpha _{jk}\right\vert
,$ required to satisfy the validity condition, with such values going to infinity with the order of the coefficient. Hence, \eqref
{largeas} is of practical value for low values of $m$, for the first Fourier coefficients. In the supplementary material, section B, we further discuss the summation of the spin chiral term, including the explicit example of the first Fourier coefficient.\\ 

The term \eqref{scalar} is not the full contribution to the coefficient of the scalar chirality term, there is also another term which is obtained by going to fourth-order perturbation theory \cite{kitamura2017probing}. It would be interesting to evaluate such term as well in the same manner. Simpler cases, such as the Fourier coefficient of the time-periodic modified exchange interaction of the Heisenberg model, after going to second order in perturbation theory (the lowest-order contribution of the electric field on the spin interaction) and before obtaining the effective static Hamiltonian \cite{mentink2015ultrafast} (see Eq. 31 in the review \cite{kitamura2020current}) are also immediately obtained with \eqref{Jeff 2} or, more generally, \eqref{general}.

\section{Outlook}

We have obtained an analytical characterization of the pivotal time reversals that have been previously investigated in the study of time-periodically modulated Mott insulators \cite{mentink2015ultrafast}. This was achieved by leveraging the existing yet hitherto overlooked summation of the pervasive summation expressions for the effective exchange interactions \cite{bukov2015universal,bukov2016schrieffer,mentink2015ultrafast}, that emerge when examining time-periodically externally modulated models, such as the Hubbard model \cite{mentink2017manipulating,oka2019floquet}. This results in a fully analytic and succinct characterization of the effective interactions, which in turn allows for asymptotic analysis, yielding significantly simplified expressions of the trigonometric type. It should be noted that the asymptotics for this regime of large amplitudes, in contrast to the case of small amplitudes, cannot be ascertained without first carrying out the summation. In conclusion, the full resummation, given the pervasiveness of the expressions in the Floquet literature, represents a relevant tool in the study of processes in Floquet physics to second order in perturbation theory.\\ 

Additionally, the straightforward nature of the resulting formula indicates that the time reversal points are simply given by Bessel zeros. These can be characterized by controlling the amplitude of the external monochromatic wave while maintaining the physical parameters of the model constant, or conversely, by tuning the Hubbard model parameter $U$. Both cases can be identified analytically, thanks to mathematical results on Bessel functions and the simplicity of the main formula \eqref{Jeff 2}. Likewise, this is also crucial in establishing the speciality and great simplicity of the model in the Floquet setting at values of the $U$ parameter halfway between resonances. The application of the formula to multi orbital models works equally well. However, it has been shown and developed in the supplementary material that extending the summation considerations to the multidimensional setting presents analytical challenges. This is particularly relevant when seeking to achieve the same degree of analytical control in the study of a number of Kitaev materials.

     

\vspace{12pt} 
\begin{acknowledgments}
We thank David Pérez-García and Jianpeng Liu for stimulating discussions and correspondence. The work was started
at Universidad Complutense de Madrid and continued and finished after moving to the Shanghai Institute for Mathematics and Interdisciplinary Sciences (SIMIS). We acknowledge previous financial support from FEI-EU-22-06, funded by Universidad Complutense de Madrid, and grant PID2020-113523GB-I00, funded by the Spanish Ministry of Science and Innovation. The work was financially supported by the  Ministry of Economic Affairs and Digital Transformation of the Spanish Government through the QUANTUM ENIA project call - Quantum Spain project, by the European Union through the Recovery, Transformation and Resilience Plan - NextGenerationEU within the framework of the Digital Spain 2026 Agenda.
\end{acknowledgments}

\bibliography{effective}

\begin{thebibliography}{62}%
\makeatletter
\providecommand \@ifxundefined [1]{%
 \@ifx{#1\undefined}
}%
\providecommand \@ifnum [1]{%
 \ifnum #1\expandafter \@firstoftwo
 \else \expandafter \@secondoftwo
 \fi
}%
\providecommand \@ifx [1]{%
 \ifx #1\expandafter \@firstoftwo
 \else \expandafter \@secondoftwo
 \fi
}%
\providecommand \natexlab [1]{#1}%
\providecommand \enquote  [1]{``#1''}%
\providecommand \bibnamefont  [1]{#1}%
\providecommand \bibfnamefont [1]{#1}%
\providecommand \citenamefont [1]{#1}%
\providecommand \href@noop [0]{\@secondoftwo}%
\providecommand \href [0]{\begingroup \@sanitize@url \@href}%
\providecommand \@href[1]{\@@startlink{#1}\@@href}%
\providecommand \@@href[1]{\endgroup#1\@@endlink}%
\providecommand \@sanitize@url [0]{\catcode `\\12\catcode `\$12\catcode `\&12\catcode `\#12\catcode `\^12\catcode `\_12\catcode `\%12\relax}%
\providecommand \@@startlink[1]{}%
\providecommand \@@endlink[0]{}%
\providecommand \url  [0]{\begingroup\@sanitize@url \@url }%
\providecommand \@url [1]{\endgroup\@href {#1}{\urlprefix }}%
\providecommand \urlprefix  [0]{URL }%
\providecommand \Eprint [0]{\href }%
\providecommand \doibase [0]{https://doi.org/}%
\providecommand \selectlanguage [0]{\@gobble}%
\providecommand \bibinfo  [0]{\@secondoftwo}%
\providecommand \bibfield  [0]{\@secondoftwo}%
\providecommand \translation [1]{[#1]}%
\providecommand \BibitemOpen [0]{}%
\providecommand \bibitemStop [0]{}%
\providecommand \bibitemNoStop [0]{.\EOS\space}%
\providecommand \EOS [0]{\spacefactor3000\relax}%
\providecommand \BibitemShut  [1]{\csname bibitem#1\endcsname}%
\let\auto@bib@innerbib\@empty
\bibitem [{\citenamefont {Bukov}\ \emph {et~al.}(2015)\citenamefont {Bukov}, \citenamefont {D'Alessio},\ and\ \citenamefont {Polkovnikov}}]{bukov2015universal}%
  \BibitemOpen
  \bibfield  {author} {\bibinfo {author} {\bibfnamefont {M.}~\bibnamefont {Bukov}}, \bibinfo {author} {\bibfnamefont {L.}~\bibnamefont {D'Alessio}},\ and\ \bibinfo {author} {\bibfnamefont {A.}~\bibnamefont {Polkovnikov}},\ }\bibfield  {title} {\bibinfo {title} {Universal high-frequency behavior of periodically driven systems: from dynamical stabilization to floquet engineering},\ }\href {https://doi.org/10.1080/00018732.2015.1055918} {\bibfield  {journal} {\bibinfo  {journal} {Advances in Physics}\ }\textbf {\bibinfo {volume} {64}},\ \bibinfo {pages} {139} (\bibinfo {year} {2015})}\BibitemShut {NoStop}%
\bibitem [{\citenamefont {Oka}\ and\ \citenamefont {Kitamura}(2019)}]{oka2019floquet}%
  \BibitemOpen
  \bibfield  {author} {\bibinfo {author} {\bibfnamefont {T.}~\bibnamefont {Oka}}\ and\ \bibinfo {author} {\bibfnamefont {S.}~\bibnamefont {Kitamura}},\ }\bibfield  {title} {\bibinfo {title} {Floquet engineering of quantum materials},\ }\href {https://doi.org/10.1146/annurev-conmatphys-031218-013423} {\bibfield  {journal} {\bibinfo  {journal} {Annual Review of Condensed Matter Physics}\ }\textbf {\bibinfo {volume} {10}},\ \bibinfo {pages} {387} (\bibinfo {year} {2019})}\BibitemShut {NoStop}%
\bibitem [{\citenamefont {de~la Torre}\ \emph {et~al.}(2021)\citenamefont {de~la Torre}, \citenamefont {Kennes}, \citenamefont {Claassen}, \citenamefont {Gerber}, \citenamefont {McIver},\ and\ \citenamefont {Sentef}}]{de2021colloquium}%
  \BibitemOpen
  \bibfield  {author} {\bibinfo {author} {\bibfnamefont {A.}~\bibnamefont {de~la Torre}}, \bibinfo {author} {\bibfnamefont {D.~M.}\ \bibnamefont {Kennes}}, \bibinfo {author} {\bibfnamefont {M.}~\bibnamefont {Claassen}}, \bibinfo {author} {\bibfnamefont {S.}~\bibnamefont {Gerber}}, \bibinfo {author} {\bibfnamefont {J.~W.}\ \bibnamefont {McIver}},\ and\ \bibinfo {author} {\bibfnamefont {M.~A.}\ \bibnamefont {Sentef}},\ }\bibfield  {title} {\bibinfo {title} {Colloquium: Nonthermal pathways to ultrafast control in quantum materials},\ }\href {https://doi.org/10.1103/RevModPhys.93.041002} {\bibfield  {journal} {\bibinfo  {journal} {Reviews of Modern Physics}\ }\textbf {\bibinfo {volume} {93}},\ \bibinfo {pages} {041002} (\bibinfo {year} {2021})}\BibitemShut {NoStop}%
\bibitem [{\citenamefont {Rudner}\ and\ \citenamefont {Lindner}(2020)}]{rudner2020band}%
  \BibitemOpen
  \bibfield  {author} {\bibinfo {author} {\bibfnamefont {M.~S.}\ \bibnamefont {Rudner}}\ and\ \bibinfo {author} {\bibfnamefont {N.~H.}\ \bibnamefont {Lindner}},\ }\bibfield  {title} {\bibinfo {title} {Band structure engineering and non-equilibrium dynamics in floquet topological insulators},\ }\href {https://doi.org/10.1038/s42254-020-0170-z} {\bibfield  {journal} {\bibinfo  {journal} {Nature reviews physics}\ }\textbf {\bibinfo {volume} {2}},\ \bibinfo {pages} {229} (\bibinfo {year} {2020})}\BibitemShut {NoStop}%
\bibitem [{\citenamefont {Floquet}(1883)}]{floquet1883equations}%
  \BibitemOpen
  \bibfield  {author} {\bibinfo {author} {\bibfnamefont {G.}~\bibnamefont {Floquet}},\ }\bibfield  {title} {\bibinfo {title} {Sur les {\'e}quations diff{\'e}rentielles lin{\'e}aires {\`a} coefficients p{\'e}riodiques},\ }in\ \href {https://doi.org/10.24033/asens.220} {\emph {\bibinfo {booktitle} {Annales scientifiques de l'{\'E}cole normale sup{\'e}rieure}}},\ Vol.~\bibinfo {volume} {12}\ (\bibinfo {year} {1883})\ pp.\ \bibinfo {pages} {47--88}\BibitemShut {NoStop}%
\bibitem [{\citenamefont {Shirley}(1965)}]{shirley1965solution}%
  \BibitemOpen
  \bibfield  {author} {\bibinfo {author} {\bibfnamefont {J.~H.}\ \bibnamefont {Shirley}},\ }\bibfield  {title} {\bibinfo {title} {Solution of the schr{\"o}dinger equation with a hamiltonian periodic in time},\ }\href {https://doi.org/10.1103/PhysRev.138.B979} {\bibfield  {journal} {\bibinfo  {journal} {Physical Review}\ }\textbf {\bibinfo {volume} {138}},\ \bibinfo {pages} {B979} (\bibinfo {year} {1965})}\BibitemShut {NoStop}%
\bibitem [{\citenamefont {Rodriguez-Vega}\ \emph {et~al.}(2021)\citenamefont {Rodriguez-Vega}, \citenamefont {Vogl},\ and\ \citenamefont {Fiete}}]{rodriguez2021low}%
  \BibitemOpen
  \bibfield  {author} {\bibinfo {author} {\bibfnamefont {M.}~\bibnamefont {Rodriguez-Vega}}, \bibinfo {author} {\bibfnamefont {M.}~\bibnamefont {Vogl}},\ and\ \bibinfo {author} {\bibfnamefont {G.~A.}\ \bibnamefont {Fiete}},\ }\bibfield  {title} {\bibinfo {title} {Low-frequency and moir{\'e}--floquet engineering: A review},\ }\href {https://doi.org/10.1016/j.aop.2021.168434} {\bibfield  {journal} {\bibinfo  {journal} {Annals of Physics}\ }\textbf {\bibinfo {volume} {435}},\ \bibinfo {pages} {168434} (\bibinfo {year} {2021})}\BibitemShut {NoStop}%
\bibitem [{\citenamefont {Bukov}\ \emph {et~al.}(2016)\citenamefont {Bukov}, \citenamefont {Kolodrubetz},\ and\ \citenamefont {Polkovnikov}}]{bukov2016schrieffer}%
  \BibitemOpen
  \bibfield  {author} {\bibinfo {author} {\bibfnamefont {M.}~\bibnamefont {Bukov}}, \bibinfo {author} {\bibfnamefont {M.}~\bibnamefont {Kolodrubetz}},\ and\ \bibinfo {author} {\bibfnamefont {A.}~\bibnamefont {Polkovnikov}},\ }\bibfield  {title} {\bibinfo {title} {Schrieffer-wolff transformation for periodically driven systems: Strongly correlated systems with artificial gauge fields},\ }\href {https://doi.org/10.1103/PhysRevLett.116.125301} {\bibfield  {journal} {\bibinfo  {journal} {Physical review letters}\ }\textbf {\bibinfo {volume} {116}},\ \bibinfo {pages} {125301} (\bibinfo {year} {2016})}\BibitemShut {NoStop}%
\bibitem [{\citenamefont {Mentink}\ \emph {et~al.}(2015)\citenamefont {Mentink}, \citenamefont {Balzer},\ and\ \citenamefont {Eckstein}}]{mentink2015ultrafast}%
  \BibitemOpen
  \bibfield  {author} {\bibinfo {author} {\bibfnamefont {J.}~\bibnamefont {Mentink}}, \bibinfo {author} {\bibfnamefont {K.}~\bibnamefont {Balzer}},\ and\ \bibinfo {author} {\bibfnamefont {M.}~\bibnamefont {Eckstein}},\ }\bibfield  {title} {\bibinfo {title} {Ultrafast and reversible control of the exchange interaction in mott insulators},\ }\href {https://doi.org/10.1038/ncomms7708} {\bibfield  {journal} {\bibinfo  {journal} {Nature communications}\ }\textbf {\bibinfo {volume} {6}},\ \bibinfo {pages} {6708} (\bibinfo {year} {2015})}\BibitemShut {NoStop}%
\bibitem [{\citenamefont {Mentink}(2017)}]{mentink2017manipulating}%
  \BibitemOpen
  \bibfield  {author} {\bibinfo {author} {\bibfnamefont {J.}~\bibnamefont {Mentink}},\ }\bibfield  {title} {\bibinfo {title} {Manipulating magnetism by ultrafast control of the exchange interaction},\ }\href {https://doi.org/10.1088/1361-648X/aa8abf} {\bibfield  {journal} {\bibinfo  {journal} {Journal of Physics: Condensed Matter}\ }\textbf {\bibinfo {volume} {29}},\ \bibinfo {pages} {453001} (\bibinfo {year} {2017})}\BibitemShut {NoStop}%
\bibitem [{\citenamefont {Hubbard}(1963)}]{hubbard1964electron}%
  \BibitemOpen
  \bibfield  {author} {\bibinfo {author} {\bibfnamefont {J.}~\bibnamefont {Hubbard}},\ }\bibfield  {title} {\bibinfo {title} {Electron correlations in narrow energy bands.},\ }\href {https://doi.org/10.1098/rspa.1963.0204} {\bibfield  {journal} {\bibinfo  {journal} {Proceedings of the Royal Society of London. Series A. Mathematical and Physical Sciences}\ }\textbf {\bibinfo {volume} {276}},\ \bibinfo {pages} {238} (\bibinfo {year} {1963})}\BibitemShut {NoStop}%
\bibitem [{\citenamefont {Arovas}\ \emph {et~al.}(2022)\citenamefont {Arovas}, \citenamefont {Berg}, \citenamefont {Kivelson},\ and\ \citenamefont {Raghu}}]{arovas2022hubbard}%
  \BibitemOpen
  \bibfield  {author} {\bibinfo {author} {\bibfnamefont {D.~P.}\ \bibnamefont {Arovas}}, \bibinfo {author} {\bibfnamefont {E.}~\bibnamefont {Berg}}, \bibinfo {author} {\bibfnamefont {S.~A.}\ \bibnamefont {Kivelson}},\ and\ \bibinfo {author} {\bibfnamefont {S.}~\bibnamefont {Raghu}},\ }\bibfield  {title} {\bibinfo {title} {The hubbard model},\ }\href {https://doi.org/10.1146/annurev-conmatphys-031620-102024} {\bibfield  {journal} {\bibinfo  {journal} {Annual review of condensed matter physics}\ }\textbf {\bibinfo {volume} {13}},\ \bibinfo {pages} {239} (\bibinfo {year} {2022})}\BibitemShut {NoStop}%
\bibitem [{\citenamefont {Peierls}(1933)}]{peierls1933theorie}%
  \BibitemOpen
  \bibfield  {author} {\bibinfo {author} {\bibfnamefont {R.}~\bibnamefont {Peierls}},\ }\bibfield  {title} {\bibinfo {title} {Zur theorie des diamagnetismus von leitungselektronen},\ }\href {https://doi.org/10.1007/BF01342591} {\bibfield  {journal} {\bibinfo  {journal} {Zeitschrift f{\"u}r Physik}\ }\textbf {\bibinfo {volume} {80}},\ \bibinfo {pages} {763} (\bibinfo {year} {1933})}\BibitemShut {NoStop}%
\bibitem [{\citenamefont {Luttinger}(1951)}]{luttinger1951effect}%
  \BibitemOpen
  \bibfield  {author} {\bibinfo {author} {\bibfnamefont {J.}~\bibnamefont {Luttinger}},\ }\bibfield  {title} {\bibinfo {title} {The effect of a magnetic field on electrons in a periodic potential},\ }\href {https://doi.org/10.1103/PhysRev.84.814} {\bibfield  {journal} {\bibinfo  {journal} {Physical Review}\ }\textbf {\bibinfo {volume} {84}},\ \bibinfo {pages} {814} (\bibinfo {year} {1951})}\BibitemShut {NoStop}%
\bibitem [{\citenamefont {Auerbach}(1998)}]{auerbach1998interacting}%
  \BibitemOpen
  \bibfield  {author} {\bibinfo {author} {\bibfnamefont {A.}~\bibnamefont {Auerbach}},\ }\href@noop {} {\emph {\bibinfo {title} {Interacting electrons and quantum magnetism}}}\ (\bibinfo  {publisher} {Springer Science \& Business Media},\ \bibinfo {year} {1998})\BibitemShut {NoStop}%
\bibitem [{\citenamefont {Kitaev}(2006)}]{kitaev2006anyons}%
  \BibitemOpen
  \bibfield  {author} {\bibinfo {author} {\bibfnamefont {A.}~\bibnamefont {Kitaev}},\ }\bibfield  {title} {\bibinfo {title} {Anyons in an exactly solved model and beyond},\ }\href {https://doi.org/10.1016/j.aop.2005.10.005} {\bibfield  {journal} {\bibinfo  {journal} {Annals of Physics}\ }\textbf {\bibinfo {volume} {321}},\ \bibinfo {pages} {2} (\bibinfo {year} {2006})}\BibitemShut {NoStop}%
\bibitem [{\citenamefont {Savary}\ and\ \citenamefont {Balents}(2016)}]{savary2016quantum}%
  \BibitemOpen
  \bibfield  {author} {\bibinfo {author} {\bibfnamefont {L.}~\bibnamefont {Savary}}\ and\ \bibinfo {author} {\bibfnamefont {L.}~\bibnamefont {Balents}},\ }\bibfield  {title} {\bibinfo {title} {Quantum spin liquids: a review},\ }\href {https://doi.org/10.1088/0034-4885/80/1/016502} {\bibfield  {journal} {\bibinfo  {journal} {Reports on Progress in Physics}\ }\textbf {\bibinfo {volume} {80}},\ \bibinfo {pages} {016502} (\bibinfo {year} {2016})}\BibitemShut {NoStop}%
\bibitem [{\citenamefont {Broholm}\ \emph {et~al.}(2020)\citenamefont {Broholm}, \citenamefont {Cava}, \citenamefont {Kivelson}, \citenamefont {Nocera}, \citenamefont {Norman},\ and\ \citenamefont {Senthil}}]{broholm2020quantum}%
  \BibitemOpen
  \bibfield  {author} {\bibinfo {author} {\bibfnamefont {C.}~\bibnamefont {Broholm}}, \bibinfo {author} {\bibfnamefont {R.}~\bibnamefont {Cava}}, \bibinfo {author} {\bibfnamefont {S.}~\bibnamefont {Kivelson}}, \bibinfo {author} {\bibfnamefont {D.}~\bibnamefont {Nocera}}, \bibinfo {author} {\bibfnamefont {M.}~\bibnamefont {Norman}},\ and\ \bibinfo {author} {\bibfnamefont {T.}~\bibnamefont {Senthil}},\ }\bibfield  {title} {\bibinfo {title} {Quantum spin liquids},\ }\href {https://doi.org/10.1126/science.aay0668} {\bibfield  {journal} {\bibinfo  {journal} {Science}\ }\textbf {\bibinfo {volume} {367}},\ \bibinfo {pages} {eaay0668} (\bibinfo {year} {2020})}\BibitemShut {NoStop}%
\bibitem [{\citenamefont {Motome}\ and\ \citenamefont {Nasu}(2020)}]{motome2020hunting}%
  \BibitemOpen
  \bibfield  {author} {\bibinfo {author} {\bibfnamefont {Y.}~\bibnamefont {Motome}}\ and\ \bibinfo {author} {\bibfnamefont {J.}~\bibnamefont {Nasu}},\ }\bibfield  {title} {\bibinfo {title} {Hunting majorana fermions in kitaev magnets},\ }\href {https://doi.org/10.7566/JPSJ.89.012002} {\bibfield  {journal} {\bibinfo  {journal} {Journal of the Physical Society of Japan}\ }\textbf {\bibinfo {volume} {89}},\ \bibinfo {pages} {012002} (\bibinfo {year} {2020})}\BibitemShut {NoStop}%
\bibitem [{\citenamefont {Hermanns}\ \emph {et~al.}(2018)\citenamefont {Hermanns}, \citenamefont {Kimchi},\ and\ \citenamefont {Knolle}}]{hermanns2018physics}%
  \BibitemOpen
  \bibfield  {author} {\bibinfo {author} {\bibfnamefont {M.}~\bibnamefont {Hermanns}}, \bibinfo {author} {\bibfnamefont {I.}~\bibnamefont {Kimchi}},\ and\ \bibinfo {author} {\bibfnamefont {J.}~\bibnamefont {Knolle}},\ }\bibfield  {title} {\bibinfo {title} {Physics of the kitaev model: fractionalization, dynamic correlations, and material connections},\ }\href {https://doi.org/10.1146/annurev-conmatphys-033117-053934} {\bibfield  {journal} {\bibinfo  {journal} {Annual Review of Condensed Matter Physics}\ }\textbf {\bibinfo {volume} {9}},\ \bibinfo {pages} {17} (\bibinfo {year} {2018})}\BibitemShut {NoStop}%
\bibitem [{\citenamefont {Trebst}\ and\ \citenamefont {Hickey}(2022)}]{trebst2022kitaev}%
  \BibitemOpen
  \bibfield  {author} {\bibinfo {author} {\bibfnamefont {S.}~\bibnamefont {Trebst}}\ and\ \bibinfo {author} {\bibfnamefont {C.}~\bibnamefont {Hickey}},\ }\bibfield  {title} {\bibinfo {title} {Kitaev materials},\ }\href {https://doi.org/10.1016/j.physrep.2021.11.003} {\bibfield  {journal} {\bibinfo  {journal} {Physics Reports}\ }\textbf {\bibinfo {volume} {950}},\ \bibinfo {pages} {1} (\bibinfo {year} {2022})}\BibitemShut {NoStop}%
\bibitem [{\citenamefont {Kitamura}\ \emph {et~al.}(2017)\citenamefont {Kitamura}, \citenamefont {Oka},\ and\ \citenamefont {Aoki}}]{kitamura2017probing}%
  \BibitemOpen
  \bibfield  {author} {\bibinfo {author} {\bibfnamefont {S.}~\bibnamefont {Kitamura}}, \bibinfo {author} {\bibfnamefont {T.}~\bibnamefont {Oka}},\ and\ \bibinfo {author} {\bibfnamefont {H.}~\bibnamefont {Aoki}},\ }\bibfield  {title} {\bibinfo {title} {Probing and controlling spin chirality in mott insulators by circularly polarized laser},\ }\href {https://doi.org/10.1103/PhysRevB.96.014406} {\bibfield  {journal} {\bibinfo  {journal} {Physical Review B}\ }\textbf {\bibinfo {volume} {96}},\ \bibinfo {pages} {014406} (\bibinfo {year} {2017})}\BibitemShut {NoStop}%
\bibitem [{\citenamefont {Itin}\ and\ \citenamefont {Katsnelson}(2015)}]{itin2015effective}%
  \BibitemOpen
  \bibfield  {author} {\bibinfo {author} {\bibfnamefont {A.}~\bibnamefont {Itin}}\ and\ \bibinfo {author} {\bibfnamefont {M.}~\bibnamefont {Katsnelson}},\ }\bibfield  {title} {\bibinfo {title} {Effective hamiltonians for rapidly driven many-body lattice systems: Induced exchange interactions and density-dependent hoppings},\ }\href {https://doi.org/10.1103/PhysRevLett.115.075301} {\bibfield  {journal} {\bibinfo  {journal} {Physical review letters}\ }\textbf {\bibinfo {volume} {115}},\ \bibinfo {pages} {075301} (\bibinfo {year} {2015})}\BibitemShut {NoStop}%
\bibitem [{\citenamefont {Hejazi}\ \emph {et~al.}(2019)\citenamefont {Hejazi}, \citenamefont {Liu},\ and\ \citenamefont {Balents}}]{hejazi2019floquet}%
  \BibitemOpen
  \bibfield  {author} {\bibinfo {author} {\bibfnamefont {K.}~\bibnamefont {Hejazi}}, \bibinfo {author} {\bibfnamefont {J.}~\bibnamefont {Liu}},\ and\ \bibinfo {author} {\bibfnamefont {L.}~\bibnamefont {Balents}},\ }\bibfield  {title} {\bibinfo {title} {Floquet spin and spin-orbital hamiltonians and doublon-holon generations in periodically driven mott insulators},\ }\href {https://doi.org/10.1103/PhysRevB.99.205111} {\bibfield  {journal} {\bibinfo  {journal} {Physical Review B}\ }\textbf {\bibinfo {volume} {99}},\ \bibinfo {pages} {205111} (\bibinfo {year} {2019})}\BibitemShut {NoStop}%
\bibitem [{Note1()}]{Note1}%
  \BibitemOpen
  \bibinfo {note} {It is widely recognised that this expression is frequently encountered in the investigation of quantum Floquet systems - and beyond.}\BibitemShut {Stop}%
\bibitem [{\citenamefont {Sen}(1952)}]{sen1952solar}%
  \BibitemOpen
  \bibfield  {author} {\bibinfo {author} {\bibfnamefont {H.~K.}\ \bibnamefont {Sen}},\ }\bibfield  {title} {\bibinfo {title} {Solar" enhanced radiation" and plasma oscillations},\ }\href {https://doi.org/10.1103/PhysRev.88.816} {\bibfield  {journal} {\bibinfo  {journal} {Physical Review}\ }\textbf {\bibinfo {volume} {88}},\ \bibinfo {pages} {816} (\bibinfo {year} {1952})}\BibitemShut {NoStop}%
\bibitem [{\citenamefont {Newberger}(1982)}]{newberger1982new}%
  \BibitemOpen
  \bibfield  {author} {\bibinfo {author} {\bibfnamefont {B.~S.}\ \bibnamefont {Newberger}},\ }\bibfield  {title} {\bibinfo {title} {New sum rule for products of bessel functions with application to plasma physics},\ }\href {https://doi.org/10.1063/1.525510} {\bibfield  {journal} {\bibinfo  {journal} {Journal of Mathematical Physics}\ }\textbf {\bibinfo {volume} {23}},\ \bibinfo {pages} {1278} (\bibinfo {year} {1982})}\BibitemShut {NoStop}%
\bibitem [{\citenamefont {Russo}\ and\ \citenamefont {Tierz}(2024)}]{russo2024landau}%
  \BibitemOpen
  \bibfield  {author} {\bibinfo {author} {\bibfnamefont {J.~G.}\ \bibnamefont {Russo}}\ and\ \bibinfo {author} {\bibfnamefont {M.}~\bibnamefont {Tierz}},\ }\bibfield  {title} {\bibinfo {title} {Landau-zener transition rates of superconducting qubits and the absorption spectrum in quantum dots},\ }\href {https://doi.org/10.1103/PhysRevA.109.033702} {\bibfield  {journal} {\bibinfo  {journal} {Physical Review A}\ }\textbf {\bibinfo {volume} {109}},\ \bibinfo {pages} {033702} (\bibinfo {year} {2024})}\BibitemShut {NoStop}%
\bibitem [{\citenamefont {Kitamura}\ \emph {et~al.}(2020)\citenamefont {Kitamura}, \citenamefont {Nagaosa},\ and\ \citenamefont {Morimoto}}]{kitamura2020current}%
  \BibitemOpen
  \bibfield  {author} {\bibinfo {author} {\bibfnamefont {S.}~\bibnamefont {Kitamura}}, \bibinfo {author} {\bibfnamefont {N.}~\bibnamefont {Nagaosa}},\ and\ \bibinfo {author} {\bibfnamefont {T.}~\bibnamefont {Morimoto}},\ }\bibfield  {title} {\bibinfo {title} {Current response of nonequilibrium steady states in the landau-zener problem: Nonequilibrium green's function approach},\ }\href {https://doi.org/10.1103/PhysRevB.102.245141} {\bibfield  {journal} {\bibinfo  {journal} {Physical Review B}\ }\textbf {\bibinfo {volume} {102}},\ \bibinfo {pages} {245141} (\bibinfo {year} {2020})}\BibitemShut {NoStop}%
\bibitem [{\citenamefont {Coulomb}(1936)}]{coulomb1936zeros}%
  \BibitemOpen
  \bibfield  {author} {\bibinfo {author} {\bibfnamefont {J.}~\bibnamefont {Coulomb}},\ }\bibfield  {title} {\bibinfo {title} {Sur les z{\'e}ros des fonctions de bessel consid{\'e}r{\'e}es comme fonction de l’ordre},\ }\href@noop {} {\bibfield  {journal} {\bibinfo  {journal} {Bull. Sci. Math}\ }\textbf {\bibinfo {volume} {60}},\ \bibinfo {pages} {297} (\bibinfo {year} {1936})}\BibitemShut {NoStop}%
\bibitem [{Note2()}]{Note2}%
  \BibitemOpen
  \bibinfo {note} {Notice how the $3/2$ case can be written also as the $1/2$ case plus extra terms subleading for large external driving.}\BibitemShut {Stop}%
\bibitem [{Note3()}]{Note3}%
  \BibitemOpen
  \bibinfo {note} {More, generally, the zeros of the cylindrical function, hence the result also applies to $J_{-\nu }(x)$ among others.}\BibitemShut {Stop}%
\bibitem [{\citenamefont {Flajolet}\ and\ \citenamefont {Schott}(1990)}]{flajolet1990non}%
  \BibitemOpen
  \bibfield  {author} {\bibinfo {author} {\bibfnamefont {P.}~\bibnamefont {Flajolet}}\ and\ \bibinfo {author} {\bibfnamefont {R.}~\bibnamefont {Schott}},\ }\bibfield  {title} {\bibinfo {title} {Non-overlapping partitions, continued fractions, bessel functions and a divergent series},\ }\href {https://doi.org/10.1016/S0195-6698(13)80025-X} {\bibfield  {journal} {\bibinfo  {journal} {European Journal of Combinatorics}\ }\textbf {\bibinfo {volume} {11}},\ \bibinfo {pages} {421} (\bibinfo {year} {1990})}\BibitemShut {NoStop}%
\bibitem [{Note4()}]{Note4}%
  \BibitemOpen
  \bibinfo {note} {Recall the symmetry of the formula. Hence, the reversals are near positive and negative integers.}\BibitemShut {Stop}%
\bibitem [{\citenamefont {Kanamori}(1959)}]{kanamori1959superexchange}%
  \BibitemOpen
  \bibfield  {author} {\bibinfo {author} {\bibfnamefont {J.}~\bibnamefont {Kanamori}},\ }\bibfield  {title} {\bibinfo {title} {Superexchange interaction and symmetry properties of electron orbitals},\ }\href {https://doi.org/10.1016/0022-3697(59)90061-7} {\bibfield  {journal} {\bibinfo  {journal} {Journal of Physics and Chemistry of Solids}\ }\textbf {\bibinfo {volume} {10}},\ \bibinfo {pages} {87} (\bibinfo {year} {1959})}\BibitemShut {NoStop}%
\bibitem [{\citenamefont {Georges}\ \emph {et~al.}(2013)\citenamefont {Georges}, \citenamefont {Medici},\ and\ \citenamefont {Mravlje}}]{georges2013strong}%
  \BibitemOpen
  \bibfield  {author} {\bibinfo {author} {\bibfnamefont {A.}~\bibnamefont {Georges}}, \bibinfo {author} {\bibfnamefont {L.~d.}\ \bibnamefont {Medici}},\ and\ \bibinfo {author} {\bibfnamefont {J.}~\bibnamefont {Mravlje}},\ }\bibfield  {title} {\bibinfo {title} {Strong correlations from hund’s coupling},\ }\href {https://doi.org/10.1146/annurev-conmatphys-020911-125045} {\bibfield  {journal} {\bibinfo  {journal} {Annu. Rev. Condens. Matter Phys.}\ }\textbf {\bibinfo {volume} {4}},\ \bibinfo {pages} {137} (\bibinfo {year} {2013})}\BibitemShut {NoStop}%
\bibitem [{\citenamefont {Liu}\ \emph {et~al.}(2018)\citenamefont {Liu}, \citenamefont {Hejazi},\ and\ \citenamefont {Balents}}]{liu2018floquet}%
  \BibitemOpen
  \bibfield  {author} {\bibinfo {author} {\bibfnamefont {J.}~\bibnamefont {Liu}}, \bibinfo {author} {\bibfnamefont {K.}~\bibnamefont {Hejazi}},\ and\ \bibinfo {author} {\bibfnamefont {L.}~\bibnamefont {Balents}},\ }\bibfield  {title} {\bibinfo {title} {Floquet engineering of multiorbital mott insulators: Applications to orthorhombic titanates},\ }\href {https://doi.org/10.1103/PhysRevLett.121.107201} {\bibfield  {journal} {\bibinfo  {journal} {Physical review letters}\ }\textbf {\bibinfo {volume} {121}},\ \bibinfo {pages} {107201} (\bibinfo {year} {2018})}\BibitemShut {NoStop}%
\bibitem [{\citenamefont {Khaliullin}(2005)}]{khaliullin2005orbital}%
  \BibitemOpen
  \bibfield  {author} {\bibinfo {author} {\bibfnamefont {G.}~\bibnamefont {Khaliullin}},\ }\bibfield  {title} {\bibinfo {title} {Orbital order and fluctuations in mott insulators},\ }\href {https://doi.org/10.1143/PTPS.160.155} {\bibfield  {journal} {\bibinfo  {journal} {Progress of Theoretical Physics Supplement}\ }\textbf {\bibinfo {volume} {160}},\ \bibinfo {pages} {155} (\bibinfo {year} {2005})}\BibitemShut {NoStop}%
\bibitem [{\citenamefont {Jackeli}\ and\ \citenamefont {Khaliullin}(2009)}]{jackeli2009mott}%
  \BibitemOpen
  \bibfield  {author} {\bibinfo {author} {\bibfnamefont {G.}~\bibnamefont {Jackeli}}\ and\ \bibinfo {author} {\bibfnamefont {G.}~\bibnamefont {Khaliullin}},\ }\bibfield  {title} {\bibinfo {title} {Mott insulators in the strong spin-orbit coupling limit: from heisenberg to a quantum compass and kitaev models},\ }\href {https://doi.org/10.1103/PhysRevLett.102.017205} {\bibfield  {journal} {\bibinfo  {journal} {Physical review letters}\ }\textbf {\bibinfo {volume} {102}},\ \bibinfo {pages} {017205} (\bibinfo {year} {2009})}\BibitemShut {NoStop}%
\bibitem [{\citenamefont {Winter}\ \emph {et~al.}(2016)\citenamefont {Winter}, \citenamefont {Li}, \citenamefont {Jeschke},\ and\ \citenamefont {Valent{\'\i}}}]{winter2016challenges}%
  \BibitemOpen
  \bibfield  {author} {\bibinfo {author} {\bibfnamefont {S.~M.}\ \bibnamefont {Winter}}, \bibinfo {author} {\bibfnamefont {Y.}~\bibnamefont {Li}}, \bibinfo {author} {\bibfnamefont {H.~O.}\ \bibnamefont {Jeschke}},\ and\ \bibinfo {author} {\bibfnamefont {R.}~\bibnamefont {Valent{\'\i}}},\ }\bibfield  {title} {\bibinfo {title} {Challenges in design of kitaev materials: Magnetic interactions from competing energy scales},\ }\href {https://doi.org/10.1103/PhysRevB.93.214431} {\bibfield  {journal} {\bibinfo  {journal} {Physical Review B}\ }\textbf {\bibinfo {volume} {93}},\ \bibinfo {pages} {214431} (\bibinfo {year} {2016})}\BibitemShut {NoStop}%
\bibitem [{\citenamefont {Kumar}\ \emph {et~al.}(2022)\citenamefont {Kumar}, \citenamefont {Banerjee},\ and\ \citenamefont {Lin}}]{kumar2022floquet}%
  \BibitemOpen
  \bibfield  {author} {\bibinfo {author} {\bibfnamefont {U.}~\bibnamefont {Kumar}}, \bibinfo {author} {\bibfnamefont {S.}~\bibnamefont {Banerjee}},\ and\ \bibinfo {author} {\bibfnamefont {S.-Z.}\ \bibnamefont {Lin}},\ }\bibfield  {title} {\bibinfo {title} {Floquet engineering of kitaev quantum magnets},\ }\href {https://doi.org/10.1038/s42005-022-00931-1} {\bibfield  {journal} {\bibinfo  {journal} {Communications Physics}\ }\textbf {\bibinfo {volume} {5}},\ \bibinfo {pages} {157} (\bibinfo {year} {2022})}\BibitemShut {NoStop}%
\bibitem [{\citenamefont {Rau}\ \emph {et~al.}(2014)\citenamefont {Rau}, \citenamefont {Lee},\ and\ \citenamefont {Kee}}]{rau2014generic}%
  \BibitemOpen
  \bibfield  {author} {\bibinfo {author} {\bibfnamefont {J.~G.}\ \bibnamefont {Rau}}, \bibinfo {author} {\bibfnamefont {E.~K.-H.}\ \bibnamefont {Lee}},\ and\ \bibinfo {author} {\bibfnamefont {H.-Y.}\ \bibnamefont {Kee}},\ }\bibfield  {title} {\bibinfo {title} {Generic spin model for the honeycomb iridates beyond the kitaev limit},\ }\href {https://doi.org/10.1103/PhysRevLett.112.077204} {\bibfield  {journal} {\bibinfo  {journal} {Physical review letters}\ }\textbf {\bibinfo {volume} {112}},\ \bibinfo {pages} {077204} (\bibinfo {year} {2014})}\BibitemShut {NoStop}%
\bibitem [{\citenamefont {Rau}\ \emph {et~al.}(2016)\citenamefont {Rau}, \citenamefont {Lee},\ and\ \citenamefont {Kee}}]{rau2016spin}%
  \BibitemOpen
  \bibfield  {author} {\bibinfo {author} {\bibfnamefont {J.~G.}\ \bibnamefont {Rau}}, \bibinfo {author} {\bibfnamefont {E.~K.-H.}\ \bibnamefont {Lee}},\ and\ \bibinfo {author} {\bibfnamefont {H.-Y.}\ \bibnamefont {Kee}},\ }\bibfield  {title} {\bibinfo {title} {Spin-orbit physics giving rise to novel phases in correlated systems: Iridates and related materials},\ }\href {https://doi.org/10.1146/annurev-conmatphys-031115-011319} {\bibfield  {journal} {\bibinfo  {journal} {Annual Review of Condensed Matter Physics}\ }\textbf {\bibinfo {volume} {7}},\ \bibinfo {pages} {195} (\bibinfo {year} {2016})}\BibitemShut {NoStop}%
\bibitem [{Note5()}]{Note5}%
  \BibitemOpen
  \bibinfo {note} {In the Appendix, we provide a discussion of the complete setup.}\BibitemShut {Stop}%
\bibitem [{\citenamefont {Chaudhary}\ \emph {et~al.}(2019)\citenamefont {Chaudhary}, \citenamefont {Hsieh},\ and\ \citenamefont {Refael}}]{chaudhary2019orbital}%
  \BibitemOpen
  \bibfield  {author} {\bibinfo {author} {\bibfnamefont {S.}~\bibnamefont {Chaudhary}}, \bibinfo {author} {\bibfnamefont {D.}~\bibnamefont {Hsieh}},\ and\ \bibinfo {author} {\bibfnamefont {G.}~\bibnamefont {Refael}},\ }\bibfield  {title} {\bibinfo {title} {Orbital floquet engineering of exchange interactions in magnetic materials},\ }\href {https://doi.org/10.1103/PhysRevB.100.220403} {\bibfield  {journal} {\bibinfo  {journal} {Physical Review B}\ }\textbf {\bibinfo {volume} {100}},\ \bibinfo {pages} {220403} (\bibinfo {year} {2019})}\BibitemShut {NoStop}%
\bibitem [{\citenamefont {Chaudhary}\ \emph {et~al.}(2020)\citenamefont {Chaudhary}, \citenamefont {Ron}, \citenamefont {Hsieh},\ and\ \citenamefont {Refael}}]{chaudhary2020controlling}%
  \BibitemOpen
  \bibfield  {author} {\bibinfo {author} {\bibfnamefont {S.}~\bibnamefont {Chaudhary}}, \bibinfo {author} {\bibfnamefont {A.}~\bibnamefont {Ron}}, \bibinfo {author} {\bibfnamefont {D.}~\bibnamefont {Hsieh}},\ and\ \bibinfo {author} {\bibfnamefont {G.}~\bibnamefont {Refael}},\ }\bibfield  {title} {\bibinfo {title} {Controlling ligand-mediated exchange interactions in periodically driven magnetic materials},\ }\href@noop {} {\bibfield  {journal} {\bibinfo  {journal} {arXiv preprint arXiv:2009.00813}\ } (\bibinfo {year} {2020})}\BibitemShut {NoStop}%
\bibitem [{\citenamefont {Sur}\ \emph {et~al.}(2022)\citenamefont {Sur}, \citenamefont {Udupa},\ and\ \citenamefont {Sen}}]{sur2022driven}%
  \BibitemOpen
  \bibfield  {author} {\bibinfo {author} {\bibfnamefont {S.}~\bibnamefont {Sur}}, \bibinfo {author} {\bibfnamefont {A.}~\bibnamefont {Udupa}},\ and\ \bibinfo {author} {\bibfnamefont {D.}~\bibnamefont {Sen}},\ }\bibfield  {title} {\bibinfo {title} {Driven hubbard model on a triangular lattice: Tunable heisenberg antiferromagnet with a chiral three-spin term},\ }\href {https://doi.org/10.1103/PhysRevB.105.054423} {\bibfield  {journal} {\bibinfo  {journal} {Physical Review B}\ }\textbf {\bibinfo {volume} {105}},\ \bibinfo {pages} {054423} (\bibinfo {year} {2022})}\BibitemShut {NoStop}%
\bibitem [{Note6()}]{Note6}%
  \BibitemOpen
  \bibinfo {note} {The opposite limit is also immediate but that one does not require the full summation and is already given in \cite {kitamura2017probing}}\BibitemShut {NoStop}%
\bibitem [{\citenamefont {Watson}(1922)}]{watson1922treatise}%
  \BibitemOpen
  \bibfield  {author} {\bibinfo {author} {\bibfnamefont {G.~N.}\ \bibnamefont {Watson}},\ }\href@noop {} {\emph {\bibinfo {title} {A treatise on the theory of Bessel functions}}},\ Vol.~\bibinfo {volume} {2}\ (\bibinfo  {publisher} {The University Press},\ \bibinfo {year} {1922})\BibitemShut {NoStop}%
\bibitem [{\citenamefont {Bremer}(2019)}]{bremer2019algorithm}%
  \BibitemOpen
  \bibfield  {author} {\bibinfo {author} {\bibfnamefont {J.}~\bibnamefont {Bremer}},\ }\bibfield  {title} {\bibinfo {title} {An algorithm for the rapid numerical evaluation of bessel functions of real orders and arguments},\ }\href {https://doi.org/10.1007/s10444-018-9613-9} {\bibfield  {journal} {\bibinfo  {journal} {Advances in Computational Mathematics}\ }\textbf {\bibinfo {volume} {45}},\ \bibinfo {pages} {173} (\bibinfo {year} {2019})}\BibitemShut {NoStop}%
\bibitem [{\citenamefont {Ciocci}\ \emph {et~al.}(1985)\citenamefont {Ciocci}, \citenamefont {Dattoli}, \citenamefont {Dipace},\ and\ \citenamefont {Torre}}]{ciocci1985simple}%
  \BibitemOpen
  \bibfield  {author} {\bibinfo {author} {\bibfnamefont {F.}~\bibnamefont {Ciocci}}, \bibinfo {author} {\bibfnamefont {G.}~\bibnamefont {Dattoli}}, \bibinfo {author} {\bibfnamefont {A.}~\bibnamefont {Dipace}},\ and\ \bibinfo {author} {\bibfnamefont {A.}~\bibnamefont {Torre}},\ }\bibfield  {title} {\bibinfo {title} {Simple derivation of new sum rules of bessel functions},\ }\href@noop {} {\bibfield  {journal} {\bibinfo  {journal} {Nuovo Cimento B;(Italy)}\ }\textbf {\bibinfo {volume} {90}} (\bibinfo {year} {1985})}\BibitemShut {NoStop}%
\bibitem [{\citenamefont {Montaldi}\ and\ \citenamefont {Zucchelli}(1988)}]{montaldi1988sum}%
  \BibitemOpen
  \bibfield  {author} {\bibinfo {author} {\bibfnamefont {E.}~\bibnamefont {Montaldi}}\ and\ \bibinfo {author} {\bibfnamefont {G.}~\bibnamefont {Zucchelli}},\ }\bibfield  {title} {\bibinfo {title} {Sum rules of special functions, revisited},\ }\href {https://doi.org/10.1007/BF02726734} {\bibfield  {journal} {\bibinfo  {journal} {Il Nuovo Cimento B (1971-1996)}\ }\textbf {\bibinfo {volume} {102}},\ \bibinfo {pages} {229} (\bibinfo {year} {1988})}\BibitemShut {NoStop}%
\bibitem [{\citenamefont {Li}\ and\ \citenamefont {Eckstein}(2020)}]{li2020manipulating}%
  \BibitemOpen
  \bibfield  {author} {\bibinfo {author} {\bibfnamefont {J.}~\bibnamefont {Li}}\ and\ \bibinfo {author} {\bibfnamefont {M.}~\bibnamefont {Eckstein}},\ }\bibfield  {title} {\bibinfo {title} {Manipulating intertwined orders in solids with quantum light},\ }\href {https://doi.org/10.1103/PhysRevLett.125.217402} {\bibfield  {journal} {\bibinfo  {journal} {Physical Review Letters}\ }\textbf {\bibinfo {volume} {125}},\ \bibinfo {pages} {217402} (\bibinfo {year} {2020})}\BibitemShut {NoStop}%
\bibitem [{\citenamefont {Weber}\ \emph {et~al.}(2023)\citenamefont {Weber}, \citenamefont {Vi{\~n}as~Bostr{\"o}m}, \citenamefont {Claassen}, \citenamefont {Rubio},\ and\ \citenamefont {Kennes}}]{weber2023cavity}%
  \BibitemOpen
  \bibfield  {author} {\bibinfo {author} {\bibfnamefont {L.}~\bibnamefont {Weber}}, \bibinfo {author} {\bibfnamefont {E.}~\bibnamefont {Vi{\~n}as~Bostr{\"o}m}}, \bibinfo {author} {\bibfnamefont {M.}~\bibnamefont {Claassen}}, \bibinfo {author} {\bibfnamefont {A.}~\bibnamefont {Rubio}},\ and\ \bibinfo {author} {\bibfnamefont {D.~M.}\ \bibnamefont {Kennes}},\ }\bibfield  {title} {\bibinfo {title} {Cavity-renormalized quantum criticality in a honeycomb bilayer antiferromagnet},\ }\href {https://doi.org/10.1038/s42005-023-01359-x} {\bibfield  {journal} {\bibinfo  {journal} {Communications Physics}\ }\textbf {\bibinfo {volume} {6}},\ \bibinfo {pages} {247} (\bibinfo {year} {2023})}\BibitemShut {NoStop}%
\bibitem [{\citenamefont {Cahill}\ and\ \citenamefont {Glauber}(1969)}]{cahill1969density}%
  \BibitemOpen
  \bibfield  {author} {\bibinfo {author} {\bibfnamefont {K.~E.}\ \bibnamefont {Cahill}}\ and\ \bibinfo {author} {\bibfnamefont {R.~J.}\ \bibnamefont {Glauber}},\ }\bibfield  {title} {\bibinfo {title} {Density operators and quasiprobability distributions},\ }\href {https://doi.org/10.1103/PhysRev.177.1882} {\bibfield  {journal} {\bibinfo  {journal} {Physical Review}\ }\textbf {\bibinfo {volume} {177}},\ \bibinfo {pages} {1882} (\bibinfo {year} {1969})}\BibitemShut {NoStop}%
\bibitem [{\citenamefont {Nieto}(1997)}]{nieto1997displaced}%
  \BibitemOpen
  \bibfield  {author} {\bibinfo {author} {\bibfnamefont {M.~M.}\ \bibnamefont {Nieto}},\ }\bibfield  {title} {\bibinfo {title} {Displaced and squeezed number states},\ }\href {https://doi.org/10.1016/S0375-9601(97)00183-7} {\bibfield  {journal} {\bibinfo  {journal} {Physics Letters A}\ }\textbf {\bibinfo {volume} {229}},\ \bibinfo {pages} {135} (\bibinfo {year} {1997})}\BibitemShut {NoStop}%
\bibitem [{\citenamefont {Wineland}\ and\ \citenamefont {Itano}(1979)}]{wineland1979laser}%
  \BibitemOpen
  \bibfield  {author} {\bibinfo {author} {\bibfnamefont {D.~J.}\ \bibnamefont {Wineland}}\ and\ \bibinfo {author} {\bibfnamefont {W.~M.}\ \bibnamefont {Itano}},\ }\bibfield  {title} {\bibinfo {title} {Laser cooling of atoms},\ }\href {https://doi.org/10.1103/PhysRevA.20.1521} {\bibfield  {journal} {\bibinfo  {journal} {Physical Review A}\ }\textbf {\bibinfo {volume} {20}},\ \bibinfo {pages} {1521} (\bibinfo {year} {1979})}\BibitemShut {NoStop}%
\bibitem [{\citenamefont {Groenewold}(1946)}]{groenewold1946principles}%
  \BibitemOpen
  \bibfield  {author} {\bibinfo {author} {\bibfnamefont {H.~J.}\ \bibnamefont {Groenewold}},\ }\href@noop {} {\emph {\bibinfo {title} {On the principles of elementary quantum mechanics}}}\ (\bibinfo  {publisher} {Springer},\ \bibinfo {year} {1946})\BibitemShut {NoStop}%
\bibitem [{\citenamefont {Chen}\ \emph {et~al.}(1989)\citenamefont {Chen}, \citenamefont {Wilczek}, \citenamefont {Witten},\ and\ \citenamefont {Halperin}}]{chen1989anyon}%
  \BibitemOpen
  \bibfield  {author} {\bibinfo {author} {\bibfnamefont {Y.-H.}\ \bibnamefont {Chen}}, \bibinfo {author} {\bibfnamefont {F.}~\bibnamefont {Wilczek}}, \bibinfo {author} {\bibfnamefont {E.}~\bibnamefont {Witten}},\ and\ \bibinfo {author} {\bibfnamefont {B.~I.}\ \bibnamefont {Halperin}},\ }\bibfield  {title} {\bibinfo {title} {On anyon superconductivity},\ }\href {https://doi.org/10.1142/S0217979289000725} {\bibfield  {journal} {\bibinfo  {journal} {International Journal of Modern Physics B}\ }\textbf {\bibinfo {volume} {3}},\ \bibinfo {pages} {1001} (\bibinfo {year} {1989})}\BibitemShut {NoStop}%
\bibitem [{\citenamefont {Simon}\ and\ \citenamefont {Halperin}(1993)}]{simon1993finite}%
  \BibitemOpen
  \bibfield  {author} {\bibinfo {author} {\bibfnamefont {S.~H.}\ \bibnamefont {Simon}}\ and\ \bibinfo {author} {\bibfnamefont {B.~I.}\ \bibnamefont {Halperin}},\ }\bibfield  {title} {\bibinfo {title} {Finite-wave-vector electromagnetic response of fractional quantized hall states},\ }\href {https://doi.org/10.1103/PhysRevB.48.17368} {\bibfield  {journal} {\bibinfo  {journal} {Physical Review B}\ }\textbf {\bibinfo {volume} {48}},\ \bibinfo {pages} {17368} (\bibinfo {year} {1993})}\BibitemShut {NoStop}%
\bibitem [{\citenamefont {Nguyen}\ and\ \citenamefont {Gromov}(2017)}]{nguyen2017exact}%
  \BibitemOpen
  \bibfield  {author} {\bibinfo {author} {\bibfnamefont {D.~X.}\ \bibnamefont {Nguyen}}\ and\ \bibinfo {author} {\bibfnamefont {A.}~\bibnamefont {Gromov}},\ }\bibfield  {title} {\bibinfo {title} {Exact electromagnetic response of landau level electrons},\ }\href {https://doi.org/10.1103/PhysRevB.95.085151} {\bibfield  {journal} {\bibinfo  {journal} {Physical Review B}\ }\textbf {\bibinfo {volume} {95}},\ \bibinfo {pages} {085151} (\bibinfo {year} {2017})}\BibitemShut {NoStop}%
\bibitem [{\citenamefont {Wang}\ \emph {et~al.}(2017)\citenamefont {Wang}, \citenamefont {Cooper}, \citenamefont {Halperin},\ and\ \citenamefont {Stern}}]{wang2017particle}%
  \BibitemOpen
  \bibfield  {author} {\bibinfo {author} {\bibfnamefont {C.}~\bibnamefont {Wang}}, \bibinfo {author} {\bibfnamefont {N.~R.}\ \bibnamefont {Cooper}}, \bibinfo {author} {\bibfnamefont {B.~I.}\ \bibnamefont {Halperin}},\ and\ \bibinfo {author} {\bibfnamefont {A.}~\bibnamefont {Stern}},\ }\bibfield  {title} {\bibinfo {title} {Particle-hole symmetry in the fermion-chern-simons and dirac descriptions of a half-filled landau level},\ }\href {https://doi.org/10.1103/PhysRevX.7.031029} {\bibfield  {journal} {\bibinfo  {journal} {Physical Review X}\ }\textbf {\bibinfo {volume} {7}},\ \bibinfo {pages} {031029} (\bibinfo {year} {2017})}\BibitemShut {NoStop}%
\end{thebibliography}%

\onecolumngrid
\newpage
\section*{Supplementary material: The summation formula, features and extensions}

The more general form of the summation formula is \cite{newberger1982new,russo2024landau}
\begin{equation}
\sum_{n=-\infty }^{\infty }\frac{\left( -1\right) ^{n}J_{\alpha +\gamma
n}\left( z\right) J_{\beta -\gamma n}\left( z\right) }{n+\mu }=\frac{\pi }{%
\sin \left( \pi \mu \right) }J_{\alpha -\gamma \mu }\left( z\right) J_{\beta
+\gamma \mu }\left( z\right) ,  \label{general}
\end{equation}%
where $\mu \in \mathbb{C}
\mathbb{Z}
$, $\alpha ,\beta ,z\in 
\mathbb{C}$,
$\gamma \in \left( 0,1\right]$, and $\Re\left( \alpha +\beta \right) >-1$. For a proof, see \cite{newberger1982new} (\cite{russo2024landau} for a simplified discussion and discussion of particular cases).

\subsection{Asymptotics}     A simple formula can be given for  large $x$, where one has the well known Bessel asymptotics \cite{watson1922treatise}
\begin{equation}
\label{asyJ}
    J_\mu(x)\approx \frac{\sqrt{2}}{\sqrt{\pi x}}\cos(x-\mu\frac{\pi}{2}-\frac{\pi}{4})\ ,\qquad \ x\gg {\rm Max}\{ 1,|\mu|^2 \}.
\end{equation}
Thus, for large $x$
\begin{equation}
\label{Wlargex}
\frac{\pi}{\sin(\pi\mu) }\, J_\mu (x) J_{-\mu} (x) \approx \frac{1}{x} \frac{\cos(\pi\mu)+\sin(2x) }{\sin(\pi\mu)} .
\end{equation}
It is straightforward to obtain subleading contributions to this expression \cite{bremer2019algorithm}, which will contain terms mixing $\mu$ and $x$. For small $x$ it is simple to find that \cite{russo2024landau}:
\begin{equation}
\label{Wlargex}
\frac{\pi}{\sin(\pi\mu) }\, J_\mu (x) J_{-\mu} (x) \approx 
\frac{1}{\mu}\left( 1+\sum\limits_{m=1}^{\infty }\frac{\left( 2m\right) !}{2^{2m}\left( m!\right) ^{2}}\frac{x^{2m}}{\left( \mu ^{2}-1\right) \left(
\mu ^{2}-2^{2}\right) ...\left( \mu ^{2}-m^{2}\right) }\right).
\end{equation}

\subsection{Scalar chirality coefficient}

In the study of emerging scalar spin chiral terms, in \cite{kitamura2017probing} they obtain an effective static Hamiltonian as a Heisenberg model that has an emergent scalar chirality term \cite{kitamura2020current}
\begin{align}
J_{\chi ,ijk}^{\text{(h)}}=& -\sum_{m=1}^{\infty
}8|t_{ij}|^{2}|t_{jk}|^{2}U^{2}\sin 2m(\theta _{ij}-\theta _{jk})  \nonumber
\\
\times & \sum_{n,l=-\infty }^{\infty }\frac{\mathcal{J}_{n+m}(\alpha _{ij})%
\mathcal{J}_{n-m}(\alpha _{ij})\mathcal{J}_{l+m}(\alpha _{jk})\mathcal{J}%
_{l-m}(\alpha _{jk})}{m\omega \lbrack U^{2}-(n+m)^{2}\omega
^{2}][U^{2}-(l+m)^{2}\omega ^{2}]}.  \label{eq:eff-chiral-1}
\end{align}
This is not the full contribution to the coefficient of the scalar chirality term, there is also the above mentioned term obtained by going to fourth-order perturbation theory \cite{kitamura2017probing}. The iterative use of the Bessel summation formula now allows us to to evaluate the above internal double summation, which sums over the $n$ and $l$ indices exactly, as follows:
\begin{eqnarray*}
&&\sum_{n,l=-\infty }^{\infty }\frac{\mathcal{J}_{n+m}(\alpha _{ij})\mathcal{%
J}_{n-m}(\alpha _{ij})\mathcal{J}_{l+m}(\alpha _{jk})\mathcal{J}%
_{l-m}(\alpha _{jk})}{m\omega \lbrack U^{2}-(n+m)^{2}\omega
^{2}][U^{2}-(l+m)^{2}\omega ^{2}]} \\
&=&\frac{\pi ^{2}}{4U^{2}\omega ^{2}\sin ^{2}\left( \frac{\pi U}{\omega }%
\right) }\left[ J_{-\frac{U}{\omega }}(\alpha _{ij})J_{\frac{U}{\omega }%
+2m}(\alpha _{ij})+J_{\frac{U}{\omega }}(\alpha _{ij})J_{-\frac{U}{\omega }%
+2m}(\alpha _{ij})\right]  \\
&&\times \left[ J_{-\frac{U}{\omega }}(\alpha _{jk})J_{\frac{U}{\omega }%
+2m}(\alpha _{jk})+J_{\frac{U}{\omega }}(\alpha _{jk})J_{-\frac{U}{\omega }%
+2m}(\alpha _{jk})\right].
\end{eqnarray*}
This analytical expression characterizes exactly all the Fourier coefficients in the Fourier series which is \eqref{eq:eff-chiral-1} and $m$ is of course the order of the harmonic. Asymptotic expressions follow immediately from this summed form. In principle the same reasoning can be applied to, say, the aforementioned other term characterizing chirality \cite{kitamura2017probing,kitamura2020current} or, certainly, the time-dependent effective interaction given in \cite{kitamura2020current}, which is also straightforward to evaluate all of its harmonics.\\

As an example of a particular result from this expression, the first Fourier coefficient for the simplest $U=1/2$ (lowest $U$ with maximal distance between resonances) would simply read:
\begin{equation*}
a_{1}(U=\frac{1}{2};\omega =1)=\frac{\pi ^{2}}{4U^{2}}F(\alpha
_{ij})F(\alpha _{jl})
\end{equation*}
where the function $F$ denotes:
\begin{equation*}
F(x)=\frac{-2x-4x\cos (2x)+(3-2x^{2})\sin (2x)}{\pi x^{3}}
\end{equation*}
All other harmonics can be found explicitly as well, containing the same two  oscillatory terms, $\cos(2x)$ and $\sin(2x)$, but the polynomial part (and hence the amplitude of such oscillations) is of higher order for higher harmonic. See Figure $7$

\begin{figure}[h] 
\includegraphics[width=0.37\textwidth]{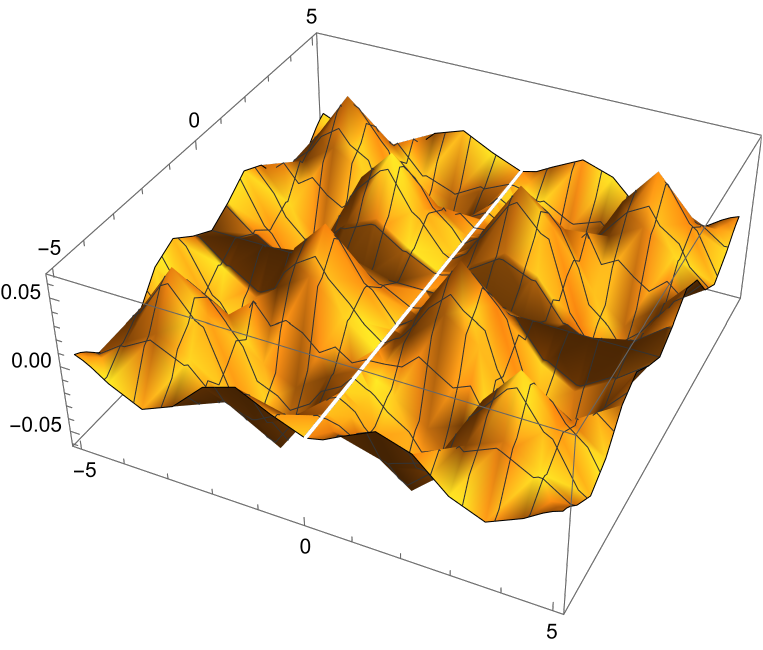}
\caption{First harmonic in terms of $\alpha
_{ij}$ and $\alpha
_{lj}$}
\end{figure}

\subsection{Extension of the formula with a phase term or time-dependent term}

It would have many physical applications if the Bessel summation formula can
be extended to a whole Fourier series or, simply put, just to add a phase
factor to it. In other words, to obtain an analytical evaluation for
\[
\sum_{n=-\infty }^{\infty }\frac{\left( -1\right) ^{n}J_{n}\left( z\right)
J_{-n}\left( z\right) }{n+\mu }e^{int}\text{ or }\sum_{n=-\infty }^{\infty }%
\frac{\left( -1\right) ^{n}J_{n}\left( z\right) J_{-n}\left( z\right) }{%
n+\mu }\cos (n\theta )
\]%
for either a time or an angle variable. Proceeding naively the same
procedure as above, one quickly obtains 
\begin{equation}
\sum_{n\in 
\mathbb{Z}
}\frac{_{J_{n}\left( z\right) ^{2}}}{n+\mu }e^{in\theta }=\frac{\pi e^{-i\mu
\theta }}{\sin \left( \pi \mu \right) }J_{\mu }\left( z\right) J_{-\mu
}\left( z\right) .  \label{e1}
\end{equation}%
or 
\begin{equation}
\sum_{n\in 
\mathbb{Z}
}\frac{_{J_{n}\left( z\right) ^{2}}}{n+\mu }\cos \left( n\theta \right) =%
\frac{\cos \left( \theta \mu \right) \pi }{\sin \left( \pi \mu \right) }%
J_{\mu }\left( z\right) J_{-\mu }\left( z\right) ,  \label{e2}
\end{equation} 
which is not correct, or it is incomplete, as can be easily checked. The
reason is because we ignored the fact that the summation formula employed
in the proof, require that the angle variable -the integration variable when using an integral representation of the Bessel function \cite{newberger1982new,russo2024landau}- satisfies 
$\phi \in \left[ -\pi ,\pi \right] $. By adding the extra
factor, now we are clearly outside this domain. 

We also can see directly that,
at least, a piece is missing and the previous expressions can not be the
whole story, by considering the moment summation formulas, which are also
useful in itself (see the discussion on Kitaev magnets, for example). In
general, a particular case of Eq. (2.17) in \cite{newberger1982new} gives%
\begin{eqnarray}
&&\sum_{n\in 
\mathbb{Z}
}n^{j}\frac{_{J_{n}\left( z\right) ^{2}}}{n+\mu }=\frac{\pi (-\mu )^{j}}{%
\sin (\pi \mu )}J_{\mu }\left( z\right) J_{-\mu }\left( z\right) 
\label{formula} \\
&&-\frac{1}{2}(-2)^{j}\sum\limits_{k=0}^{j-1}\left( \frac{\mu }{2}\right)
^{k}\left( -1\right) ^{\frac{j-1-k}{2}}\frac{d^{j-1-k}}{dx^{j-1-k}}\left[
J_{0}(2z\cos \left( \frac{x}{4}\right) \right] _{x=2\pi }  \nonumber
\end{eqnarray}%
Useful particular cases \cite{newberger1982new}: 
\begin{eqnarray}
\sum_{n\in 
\mathbb{Z}
}n\frac{_{J_{n}\left( z\right) ^{2}}}{n+\mu } &=&\frac{-\mu \pi }{\sin
\left( \pi \mu \right) }J_{\mu }\left( z\right) J_{-\mu }\left( z\right) +1
\label{moments} \\
\sum_{n\in 
\mathbb{Z}
}n^{2}\frac{_{J_{n}\left( z\right) ^{2}}}{n+\mu } &=&\frac{\mu ^{2}\pi }{%
\sin \left( \pi \mu \right) }J_{\mu }\left( z\right) J_{-\mu }\left(
z\right) -\mu   \nonumber \\
\sum_{n\in 
\mathbb{Z}
}n^{3}\frac{_{J_{n}\left( z\right) ^{2}}}{n+\mu } &=&\frac{-\mu ^{3}\pi }{%
\sin \left( \pi \mu \right) }J_{\mu }\left( z\right) J_{-\mu }\left(
z\right) +\mu ^{2}+\frac{z^{2}}{2}  \nonumber
\end{eqnarray}%
We see that the
whole summation of the first term in the moment formulas above, gives the
term that we have naively obtained. One could then consider this whole term while incorporate the remaining extra term perturbatively,
staying at low-order, using \eqref{moments}. The
moment formulas corresponding to the more general setting \eqref{general} can
be obtained from analyzing Eq. (2.17) in \cite{newberger1982new}. Therefore, for small $\theta $ one can consider the approximation:
\[
\sum_{n\in\mathbb{Z}
}\frac{_{J_{n}\left( z\right) ^{2}}}{n+\mu }\cos \left( n\theta \right)
\approx \frac{\cos \left( \theta \mu \right) \pi }{\sin \left( \pi \mu
\right) }J_{\mu }\left( z\right) J_{-\mu }\left( z\right) -\frac{\mu \theta
^{2}}{2}
\]
In \cite{ciocci1985simple} they studied
\[
\sum_{n=-\infty }^{\infty }\frac{t^{n}J_{n}^{2}\left( x\right) }{\mu +n}
\]%
By multiplying with $t^{\mu }$ and taking a parametric derivative, they
could use%
\[
\sum_{n=-\infty }^{\infty }t^{n}J_{n}^{2}\left( x\right) =I_{0}(x\frac{t-1}{%
\sqrt{t}}),\text{ \ \ \ \ }0<\left\vert t\right\vert <\infty .
\]%
With $I_{0}$ the modified Bessel function of second order. Of course,
writing $t=\exp \left( i\theta \right) $ we could have written $J_{0}(2x\sin
\theta )$ instead. It is then found%
\[
\sum_{n=-\infty }^{\infty }\frac{t^{n}J_{n}^{2}\left( x\right) }{\mu +n}%
=t^{-\mu }\int\limits_{1}^{t}t^{\prime (\mu -1)}I_{0}(x\frac{t^{\prime }-1}{%
\sqrt{t^{\prime }}})dt^{\prime }+t^{-\mu }\sum_{n=-\infty }^{\infty }\frac{%
J_{n}^{2}\left( x\right) }{\mu +n},
\]%
which is of the same form as above. It seems one can only use the power
series of $I_{0}$ and integrate term by term and this should coincide with
what we discussed above. Alternatively, expanding the term $e^{int}$ and \eqref{moments}, we obtain:
\[
\frac{\pi }{\sin \pi \mu }J_{\mu }(x)J_{-\mu }(x)e^{-it\mu }+(1-it-\mu \frac{%
t^{2}}{2}+\frac{it^{3}}{6}\left( \mu ^{2}+\frac{x^{2}}{2}\right) +...)
\]

\subsection{The case of different variables}
We have seen that the consideration of two different variables in an extended
summation 
\begin{equation}
  \sum_{n=-\infty }^{\infty }\frac{\left( -1\right) ^{n}J_{n}\left( z\right)
J_{-n}\left( w\right) }{n+\mu }, 
\label{different}  
\end{equation}
is interesting from the physical point of view, when studying the Kitaev quantum magnet or when discussing spin chiral terms. There is no formula analogous to \eqref{general} in this case. A naive conjecture, in extending naively (\eqref{general}), would be%
\[
\sum_{n=-\infty }^{\infty }\frac{\left( -1\right) ^{n}J_{n}\left( z\right)
J_{-n}\left( w\right) }{n+\mu }\approx \frac{\pi }{2\sin \left( \pi \mu
\right) }\left( J_{-\mu }\left( z\right) J_{\mu }\left( w\right) +J_{-\mu
}\left( w\right) J_{\mu }\left( z\right) \right) . 
\]%
This is not correct, but we have numerically tested the approximation and found it to be reasonably good even for values of $w$ that are not as close to $z$ for a rather large range of $\mu$.
As an oscillatory function, it is expected that at some specific points, the approximation is not as good, and we have verified this numerically. 
In fact \eqref{different} admits a remarkably different evaluation, which involves the infinite summation of a single Bessel function \cite{montaldi1988sum}: 

\begin{eqnarray*}
\sum_{n=-\infty }^{\infty }\frac{\left( -1\right) ^{n}J_{n}\left( x\right)
J_{-n}\left( y\right) }{\mu -in} &=&\frac{1}{\sinh 2\pi \mu }\int_{0}^{2\pi
}d\theta \exp \left[ i(x-y)\cos \theta \right] \mathbf{J}_{2i\mu }\left( 2%
\sqrt{xy}\sin \theta \right)  \\
&=&\frac{1}{\mu }\sum_{n=0}^{\infty }(-1)^{n}\frac{\left( 1/2\right) _{n}}{%
(1-i\mu )_{n}(1+i\mu )_{n}}\left( \frac{2xy}{x-y}\right) ^{n}J_{n}(x-y),
\end{eqnarray*}
where $\mathbf{J}_{2i\mu }$ denotes the Anger function \cite{montaldi1988sum} and $()_{n}$ is the usual Pochhammer symbol (rising factorial).

We can use this expression to approximate for the limit where the two
variables are close, which is a physically relevant setting. For $%
x\rightarrow 0$ at leading order $J_{n}(x-y)\sim \frac{\left( x-y\right) ^{n}%
}{2^{n}n!}.$ Therefore, we have, for $x,y$ close in value%
\begin{eqnarray*}
\sum_{n=-\infty }^{\infty }\frac{\left( -1\right) ^{n}J_{n}\left( x\right)
J_{-n}\left( y\right) }{\mu -in} &\simeq &\frac{1}{\mu }\sum_{n=0}^{\infty
}(-1)^{n}\frac{\left( 1/2\right) _{n}}{n!(1-i\mu )_{n}(1+i\mu )_{n}}\left(
xy\right) ^{n} \\
&=&\frac{1}{\mu }{}_1F_{2}(\frac{1}{2};1-i\mu ,1+i\mu ;-xy).
\end{eqnarray*}

Interestingly, for the known case $x=y$, which is the obvious particular case of the formula \eqref{general} and corresponds to the starting but also central result in this paper, namely that \eqref{Jeff 1}=\eqref{Jeff 2}, this hypergeometric summation may be exact because it involves an infinite sum of vanishingly small contributions, given by and then the substitution $J_{n}(x-y)= \frac{\left( x-y\right) ^{n}%
}{2^{n}n!}$ is exact. So, there is total convergence. This is indeed the case, as it can be checked numerically and hence it holds that
\begin{eqnarray*}
\sum_{n=-\infty }^{\infty }\frac{\left( -1\right) ^{n}J_{n}\left( x\right)
J_{-n}\left( x\right) }{\mu +n}&=&\frac{1}{i\mu }{}_1F_{2}(\frac{1}{2};1-i\mu ,1+i\mu ;-xy). \label{hyper}
\end{eqnarray*}

This hypergeometric interpretation could be used for an alternative or complementary study of the reversals (zeroes). A direct study of asymptotic behavior using the hypergeometric directly, gives the same result used in the paper \eqref{Wlargex}.

\subsection{The half-resonance point}

To somewhat complement the discussion of the specialness at and around the
half-resonance point, discussed in the main text, to recall, following the description in the introduction of \cite{bremer2019algorithm} that the
scaled Bessel functions $J_{\nu }(t)\sqrt{t}$ and $Y_{\nu }(t)\sqrt{t}$
satisfy the second order linear ordinary differential equation 
\begin{equation}
y^{\prime \prime }(t)+\left( 1-\frac{\nu ^{2}-\frac{1}{4}}{t^{2}}\right)
y(t)=0\ \ \mbox{for all}\ \ 0<t<\infty .  \label{introduction:besseleq}
\end{equation}%
When $0\leq \nu \leq 1/2$, the coefficient of $y$ in (\eqref%
{introduction:besseleq}) is positive on the entire half-line $(0,\infty )$,
but it is negative on the interval 
\begin{equation}
\left( 0,\sqrt{\nu ^{2}-\frac{1}{4}}\right)   \label{introduction:interval1}
\end{equation}%
and positive on 
\begin{equation}
\left( \sqrt{\nu ^{2}-\frac{1}{4}},\infty \right) 
\label{introduction:interval2}
\end{equation}%
when $\nu >1/2$. Usual WKB arguments imply that solutions of (\eqref{introduction:besseleq}) approximately behave
like increasing or decreasing exponential on (\eqref{introduction:interval1})
while are oscillatory on (\eqref{introduction:interval2}). We define
the subset 
\begin{equation}
\mathcal{O}=\left\{ (\nu ,t):0\leq \nu \leq \frac{1}{2}\ \ \mbox{and}\
t>0\right\} \bigcup \left\{ (\nu ,t):\nu >\frac{1}{2}\ \ \mbox{and}\ t\geq 
\sqrt{\nu ^{2}-\frac{1}{4}}\right\}   \label{introduction:oscillatory}
\end{equation}%
of $\mathbb{R}\times \mathbb{R}$ as the oscillatory region and the subset 
\begin{equation}
\mathcal{N}=\left\{ (\nu ,t):\nu >\frac{1}{2}\ \ \mbox{and}\ \ 0<t<\sqrt{\nu
^{2}-\frac{1}{4}}\right\}   \label{introduction:nonoscillatory}
\end{equation}%
of $\mathbb{R}\times \mathbb{R}$ as the nonoscillatory region. Let us show,
not just for a single Bessel function but for the whole formula, with a
figure, a non-oscillatory region choosing for that purpose a relatively
large value of the rescaled Hubbard parameter $U$ Indeed, we can see that the
expression for the effective interaction only starts to oscillate when the
absolute value of the external drive is roughly the value of the rescaled $U$
parameter, which in this case is $10.5$.   

\begin{figure}[h] 
\includegraphics[width=0.37\textwidth]{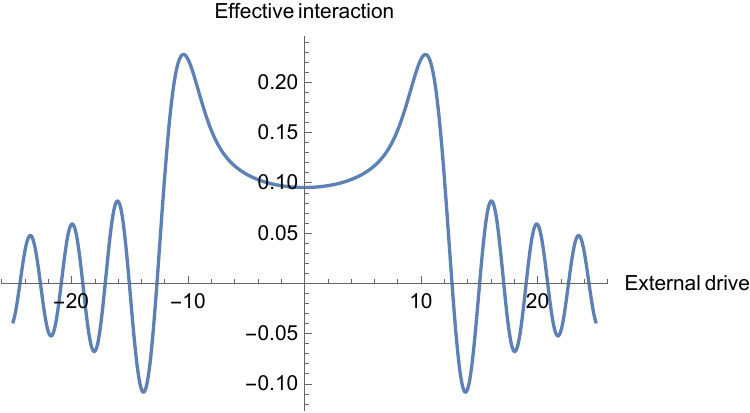}
\caption{Non-oscillatory region of effective J for $\mu=10.5$}
\end{figure}


\section{Comment on the quantum treatment of light}

A more intricate approach is necessary for a quantum analysis of light, however, the Peierls phase method remains applicable. The works \cite{li2020manipulating,weber2023cavity} have introduced the average of the displacement operator in the use of the Peierls phase method, but did not recognised it as such. The sole observation made here is that, upon realizing that it is the long-known average of the displacement operator \cite{cahill1969density}, a connection with the physics of Landau levels arises.

Recall that, in terms of the creation and annihilation operators $a=\lambda
/2+d/d\lambda $ and $a^{\dagger }=$ $\lambda /2-d/d\lambda $ the average of
the displacement operator \cite{cahill1969density} can be expressed as 
\[
D(y)\equiv \exp (y(a+a^{\dagger })).
\]%
The displacement operator acting on the vacuum state generates a coherent
state $D(y)\left\vert 0\right\rangle =\left\vert y\right\rangle $ and when
it acts on an excited state $D(y)\left\vert n\right\rangle $ of the harmonic
oscillator, these are displaced number states \cite{nieto1997displaced}, which are
also wavepackets that keep their shape and follow classical motion. 

The
evaluation of the matrix element of the displacement operator between two
harmonic oscillator states, which is the average considered in \cite{li2020manipulating,weber2023cavity} to account
for a quantum treatment of the driving light field, is a result in the
theory of coherent states and quantum optics \cite{cahill1969density} and in
laser cooling \cite{wineland1979laser} -it also appeared previously in relationship
with Wigner functions \cite{groenewold1946principles}-. Its evaluation is  \cite{cahill1969density,wineland1979laser}%
\begin{equation}
j_{n^{\prime },n}=\left\langle n^{\prime }\left\vert e^{y(a+a^{\dagger
})}\right\vert n\right\rangle =e^{-y^{2}/2}y^{\Delta n}\left( \sqrt{\frac{%
n_{<}!}{n_{>}!}}\right) L_{n_{<}}^{\Delta n}(-y^{2}),  \label{matrix}
\end{equation}
where $n_{<}=\min (n,n^{\prime })$, $n_{>}=\max (n,n^{\prime })$ and $\Delta
n=n_{>}-n_{<}$. Notice the analytical form of Landau levels appearing in the
r.h.s. of \eqref{matrix}. The effective
exchange interaction then is of the form \eqref{Jeff 1} with the replacing of the Bessel
functions \cite{li2020manipulating}
\[
J_{ex}=\sum\limits_{l\geq -n}\frac{\left( -1\right) ^{l}j_{n,n+l}j_{n+l,n}}{%
1+l\omega /U}
\]%
Making this straightforward identification provided by \eqref%
{matrix} shows that the expression for the effective exchange coupling is related to the electromagnetic
response in a Landau level system.

More precisely, it is exactly $\Sigma _{0}$, the first component of the electromagnetic response tensor in \cite{chen1989anyon}. If a
complete summation could be conducted, in addition to the quantum treatment of light, it would also extend
known results for the electromagnetic response in quantum Hall effect systems beyond the random phase approximation framework, studied in \cite{simon1993finite} and further discussed in \cite{nguyen2017exact,wang2017particle}.

\end{document}